\documentclass[reprint,amsmath,amssymb,aps]{revtex4-2}
\usepackage{graphicx}% Include figure files
\usepackage{dcolumn}% Align table columns on decimal point
\usepackage{bm}% bold math

\usepackage{amssymb}
\usepackage{amsmath}
\usepackage{epsfig}
\usepackage{chemarr}
\usepackage{url}
\usepackage{natbib}
\usepackage{color}

\usepackage[final]{changes}
\begin{document}
	
	\title{The Physical Effects of Learning}
	
	\author{Menachem Stern$^1$, Andrea J. Liu$^{1,2}$ and Vijay Balasubramanian$^{1,3,4}$}
	
	\affiliation{$^1$Department of Physics and Astronomy, University of Pennsylvania, Philadelphia, PA 19104}
	\affiliation{$^2$Center for Computational Biology, Flatiron Institute, Simons Foundation, New York, NY 10010, USA}
	\affiliation{$^3$Santa Fe Institute, 1399 Hyde Park Road, Santa Fe, NM 87501, USA}
	\affiliation{$^4$Theoretische Natuurkunde, Vrije Universiteit Brussel, Pleinlaan 2, B-1050 Brussels, Belgium}

	%\affiliation{${}^2$ Department of Mechanical Engineering, Technion, Haifa, 32000, Israel}
	
	%\affiliation{${}^3$ Department of Physics, Boston University, Boston, MA 02139}
	
	%\affiliation{${}^4$ Physics Department and the James Franck Institute,	University of Chicago, Chicago, IL 60637}

	\date{\today}
	
	\begin{abstract}
		Interacting many-body physical systems ranging from neural networks in the brain to folding proteins to self-modifying electrical circuits can learn to perform diverse tasks.  This learning, both in nature and in engineered systems, can occur through evolutionary selection or through dynamical rules that drive active learning from experience. Here, we show that \added{learning in linear physical networks with weak input signals} leaves architectural imprints on the Hessian of a physical system. Compared to a generic organization of the system components,  (a) the effective physical dimension of the response to inputs decreases, (b) the response of physical degrees of freedom to random perturbations (or system ``susceptibility'')  increases, and (c) the low-eigenvalue eigenvectors of the Hessian align with the task. Overall, these effects embody the typical scenario for learning processes in physical systems in the weak input regime, suggesting ways of discovering whether a physical network may have been trained.
		%Overall, these effects suggest a method for discovering the task that a physical network may have been trained for.
		%(the participation ratio of low-eigenvalue modes)
	\end{abstract}
	
	\pacs{Valid PACS appear here}% PACS, the Physics and Astronomy
	% Classification Scheme
	\maketitle
	
	\section{Introduction}
	Nature is replete with systems that learn. For example, animals learn new behaviors, the immune systems of vertebrates and bacteria learn pathogenic environments, and, over evolutionary time, proteins learn structures that fold just right to achieve precise and specific molecular functions. In supervised computational machine learning~\cite{lecun2015deep,mehta2019high}, the learning process is formulated as minimization of a {\it cost function} on a system's input-output behavior. More generally, this formulation provides a powerful paradigm for solving difficult inverse problems~\cite{de2005learning,ye2019optimization}. The same paradigm has also been exploited to describe biological learning in various forms,  e.g., in neural~\cite{richards2019deep} and immune systems~\cite{farmer1986immune}.
	
	Biological systems are necessarily physical in nature and therefore, like any physical system, must obey certain constraints that cause their learning processes to differ from those of computers. 
	%In particular, typical computational algorithms use processors 
	In particular, while computer algorithms often seek to globally descend cost function gradients with respect to learning degrees of freedom (e.g., neural network weights), physical systems without external processors cannot generally implement such optimization processes, even though learning by natural selection can sometimes be cast in this way.
	%in terms of global descent algorithms, general physical systems without external processors cannot directly implement such optimization processes.
	Typical biological learning, e.g., by an animal learning a new behavior,  operates on time scales far shorter than evolutionary ones, and must proceed by dynamical processes (learning rules) that modify internal (learning) degrees of freedom in response to examples~\cite{stern2023learning}. Such learning rules are generally local in space and time and cannot be informed about the functionality of the whole system. In other words, learning in physical systems on shorter-than-evolutionary time scales differs from computational machine learning in that the learning is \emph{emergent}. It is a collective behavior of many elementary units, each implementing simple rules based on its own local environment. 
	
	The Hebb rule in neuroscience (``neurons that fire together, wire together'') is an example of a local rule -- synaptic plasticity is based on local information, leading to debates about whether and how such dynamics propagate information about the training task to individual neurons and synapses.  Local rules have also been exploited to train laboratory non-biological mechanical networks to exhibit auxetic behavior~\cite{pashine2019directed,pashine2021local} or  protein-inspired functions~\cite{hexner2019effect, hexner2019periodic,stern2021physical}. Other local rules have been proposed for associative memory~\cite{stern2018shaping, markovic2020physics, grollier2020neuromorphic, stern2020continual, yu2021hopfield}; one of them has even been demonstrated in the lab~\cite{arinze2023learning}.  
	Here we will focus on a powerful set of local rules based on the framework of Contrastive Hebbian Learning, which perform approximate gradient descent of a cost function~\cite{movellan1991contrastive,scellier2017equilibrium,stern2021supervised,lopez2023self, kendall2020training, scellier2021deep, MARTIN2021102222, stern2022physical, anisetti2023learning, anisetti2022frequency}.  Such contrastive learning was recently realized experimentally for tasks including regression and classification in electronic resistor networks~\cite{dillavou2022demonstration, wycoff2022learning,dillavou2023machine} and motion tasks in elastic spring networks~\cite{altman2023experimental}.

	Here, we  focus on physical systems such as athermal mechanical, flow or electrical networks, in which the learning rate is  slow compared to the rate of physical relaxation. In this limit,  mechanical networks remain in equilibrium during the learning process, while  flow/electrical networks remain in steady state. As a result, forces on nodes of a mechanical network must add to zero, while all the currents through nodes of a flow or electrical network must add to zero. These constraints arise because the system must typically be at a minimum of a \emph{physical} cost function (e.g., the energy in a mechanical network or the dissipated power in a flow/electrical network) with respect to the physical degrees of freedom (e.g.,  node positions   in mechanical networks, or node pressures/voltages in flow/electrical networks). These physical degrees of freedom couple to the learning degrees of freedom (e.g.,  spring constants in mechanical networks or conductances for each edge in flow/electrical networks). Note that systems that operate at a minimum of  a physical cost function are naturally recurrent in the sense that information flows in all directions, not only from `inputs' to `outputs.'   Thus, they differ fundamentally from computational learning algorithms based on optimizing a feed-forward functional map.

	%Here, we show in detail how the energy landscapes of generic physical systems are restructured \emph{by physical learning}. We discuss how physical learning rules are derived in linear physical networks, relate these rules to the geometry of the energy landscape, and describe the scenario of softening low-energy modes and their alignment to the trained task. This modification of the energy landscape in the vicinity of the equilibrium state leads to ``soft'' dynamics with strong responses to generic inputs, as well as low-dimensional dynamics, such that physical degrees of freedom become correlated with each other and with learned tasks. This alignment with the learned task suggests that by studying how a trained natural network responds to random inputs, we can discover what it was trained for. In a sense, things become what they learn.  
	
	In such equilibrium or steady-state systems, learning involves a \emph{double} minimization -- the learning cost function is minimized, %with respect to the learning degrees of freedom, 
	and the physical cost function must also be minimized with respect to the physical degrees of freedom. Therefore there are two landscapes of interest: the learning cost function in the high-dimensional space spanned by the learning degrees of freedom, and the physical cost function in the high-dimensional space spanned by the physical degrees of freedom. These two landscapes are coupled together. As a system learns, either directly by double gradient descent or by using local rules that approximate gradient descent in the learning landscape, changes in the learning degrees of freedom sculpt the physical landscape. We examine the effects of learning on this physical landscape and the properties of the physical system. 
	
	%We further show this creation of soft aligned mode generalizes to training of multiple tasks on the same physical system. However, typical learning systems have a finite learning capacity, i.e., a maximum number of tasks that can be learned well, which scales with the system size~\cite{rocks2019limits,ruiz2019tuning}. We show that beyond capacity, learning makes physical systems stiff rather than soft (i.e., responses to typical inputs are small), although they remain low dimensional. The physical responses of systems during training can thus be used to assess the quality of the learning process and its prospects of success. 

	We are inspired by recent efforts in the fields of protein allostery and computational neuroscience that suggest learning has a set of typical effects on the physical structure of a network and its response to external signals. Many folded proteins, often modeled as mechanical networks~\cite{tirion1996large, yang2009protein},  have evolved allosteric function, where binding of a regulatory molecule at a ``source" site triggers a conformational change in the protein that either enables or inhibits binding of another molecule at a distant ``target" site. Model networks, as well as networks derived from folded proteins, that display allosteric behavior have been shown to exhibit low-dimensional responses to strains applied at the source~\cite{leo2005analysis, mitchell2016strain, tlusty2017physical, flechsig2017design, rocks2017designing, yan2017architecture, yan2018principles,ravasio2019mechanics, wodak2019allostery, hexner2021adaptable}. In particular, the allosteric response is well-captured by low-energy (soft) normal modes of vibration~\cite{yan2017architecture, yan2018principles, rouviere2021emergence, bhaumik2022loss, bhaumik2023mechanical, anisetti2023emergent}. It has been suggested that these soft modes are related to \textit{global epistasis}~\cite{otwinowski2018inferring}, the notion the function of proteins is encoded in low-dimensional manifolds in the space of genotypes~\cite{husain2020physical}. Similar observations have been made in neural circuits and model systems~\cite{williams2018unsupervised,yang2019task,cueva2020low,recanatesi2021predictive, jazayeri2021interpreting}, in particular for unsupervised learning tasks such as predictive coding~\cite{huang2011predictive}, a phenomenon sometimes summarized as the ``neural manifold hypothesis''.  Our goal is to understand how such effects arise in physical networks trained to perform functions.
	
	In this paper, we study the effects of learning on the physical landscapes of mechanical and flow networks, as they use local rules to learn tasks by effectively performing double gradient descent in the physical and learning landscapes. We show in detail how the structure of the physical landscape changes to accommodate the learned tasks, how it develops features observed in protein allostery studies~\cite{yan2017architecture,yan2018principles,husain2020physical,rouviere2021emergence}, and how it is affected by loading multiple concurrent tasks on the network, up to and beyond its capacity~\cite{rocks2019limits}. We find that, generically, when physical networks learn tasks in the \textit{linear response} regime, they become soft as learning proceeds, and the responses become low-dimensional, aligning with directions of low curvature in the physical landscape. The tasks that can be learned in this way include not only allostery, as in proteins, but typical computer science tasks such as regression and classification. In a sense, we are showing that {\it things become what they learn}. The implication is that learning imprints signatures on a physical system, allowing an external observer to gain insight into whether a network has been trained, and for what tasks. %We hope this work will encourage further discussion on the physical effects of learning, and how these effects can inform us about the effectiveness of a learning system, or its very ability to learn.

	The structure of the paper is as follows: In section II, we detail the physical learning process, showing how physical networks can learn to implement desired tasks. In Section III, we discuss how such learning affects the physical Hessian of the learning system and its eigenspace, leading to eigenmode alignment, increased effective conductance, and lower effective dimension. In Section IV, we look at the effects of learning multiple tasks at the same time, below and above the system capacity.

	\begin{figure}
		\includegraphics[width=0.95\linewidth]{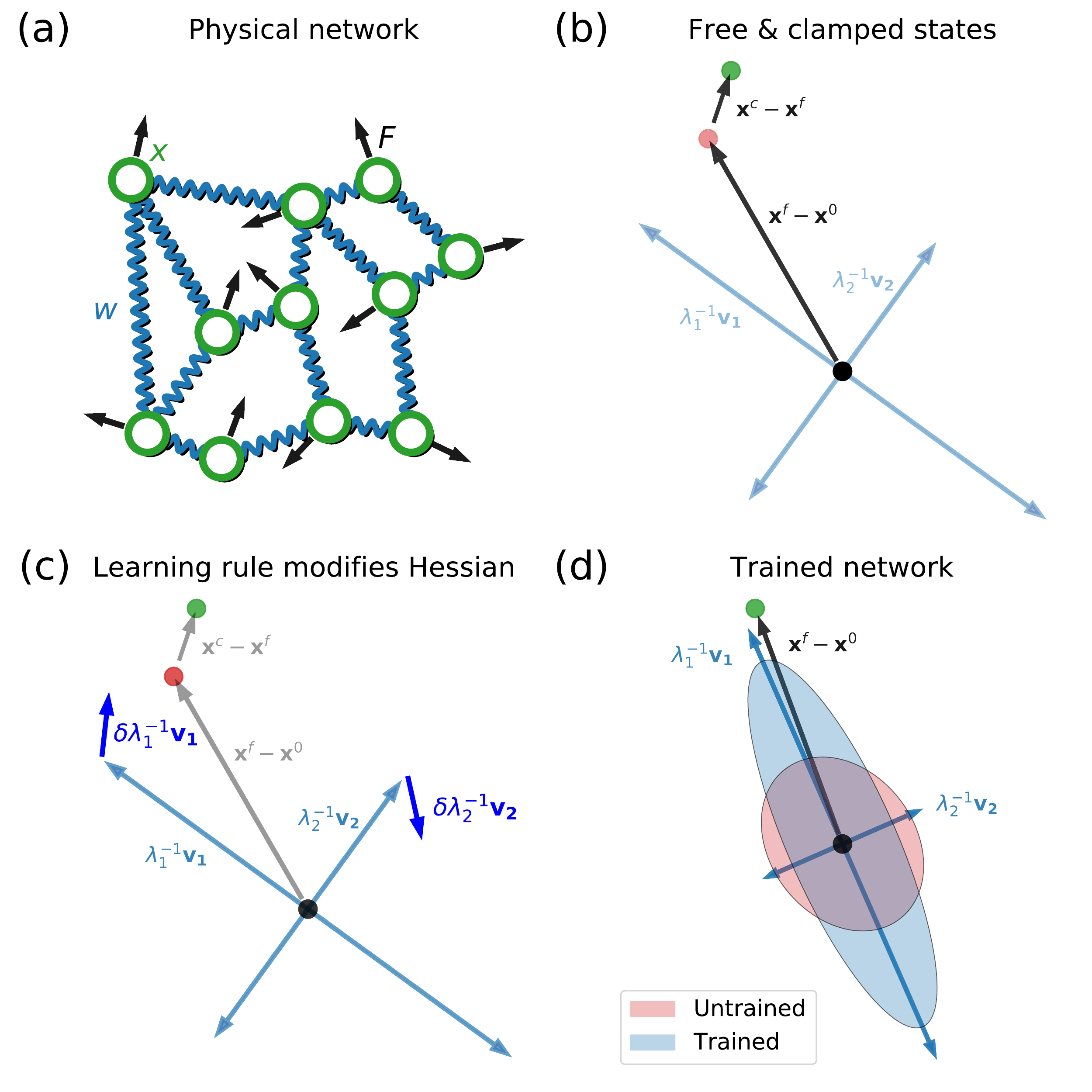}
		\caption{Learning modifies the physical network. a) Input forces (black) are applied to the physical degrees of freedom (green, e.g. node positions) of a physical network (e.g. mechanical spring network), whose interactions correspond to learning degrees of freedom (blue, e.g. spring constants). b) In the physical configuration space, this input force causes the system to respond, equilibrating in a \emph{free state} (red dot). To train the system, a further `output force' is applied, nudging the system to a \emph{clamped state} (green dot). The blue arrows describe the inherent physical coordinate system, with direction corresponding to eigenmodes $v_{ab}$ of the physical Hessian $H_{ab}$, and lengths correspond to the associated inverse eigenvalues $\lambda_{a}^{-1}$. c) A local learning rule is applied, modifying the learning degrees of freedom. On top of improving the system free state response, learning tends to rotate the Hessian coordinate system such that the eigenmode corresponding to the lower eigenvalues align with the free state response, and decrease these eigenvalues. d) Training results in a physical system whose lower eigenvalues are reduced, and eigenmodes aligned with the trained task(s). The system responds considerably more strongly to random forces, shown by the area spanned by the trained inverse eigenvalues (blue ellipse) compared to the untrained ones (red ellipse). Training makes the physical system more conductive and lower dimensional.
			\label{fig:Schematic}}
	\end{figure}

	\section{Learning in a physical network}

	Consider a network (Fig.~\ref{fig:Schematic}a) with physical degrees of freedom $x_a, \,  a=1\cdots N$ collected into a vector $\vec{x}$,  and learning degrees of freedom $w_i, \, i=1\cdots N_w$ collected into a vector $\vec{w}$. We apply inputs to one subset of the physical degrees of freedom, and designate another subset as outputs. Learning modifies the learning degrees of freedom to improve the physical responses of the outputs, driving them closer to the desired responses. For an athermal mechanical spring network, we will designate the positions of the nodes as physical degrees of freedom $\{ x_a \}$ that adjust to minimize the elastic energy, which is the physical cost function $E$. The learning degrees of freedom $w_i$ are spring stiffnesses~\cite{rocks2019limits, hexner2019effect}. The inputs are forces on nodes and outputs are node displacements. For a flow network, $\{x_a\}$ represent the node pressures, which adjust to minimize the power dissipated, which is the physical cost function $E$. The learning degrees of freedom $\{ w_i \}$ are the conductances of the edges of the network while the inputs are externally applied currents at the nodes~\cite{rocks2019limits} and outputs are node pressures.  In an electric resistor network, $\{ x_a\}$ may correspond to the node voltage values while $\{ w_i \}$ correspond to the edge resistances~\cite{kendall2020training,dillavou2022demonstration}.
	
	%The physical degrees of freedom, e.g., values at network nodes, respond to inputs we apply, while learning modifies $\vec{w}$, e.g., edge weights controlling network dynamics, to drive physical responses closer to the desired output. While the distinction between the physical and learning degrees of freedom is somewhat arbitrary, it is sensible to choose slow degrees of freedom, which may be easily controlled, as the learning variables. For example, in an electric resistor network (or an analogous flow network), the node voltages  are natural physical degrees of freedom, while  edge resistances function as the learning degrees of freedom~\cite{kendall2020training,dillavou2022demonstration}. In mechanical spring networks, the physical degrees of freedom may be node positions, or edge strains, while the edge stiffness values or rest lengths serve as learning degrees of freedom~\cite{pashine2019directed, hexner2019effect}.

	Protocols for deriving physically realizable learning rules for such systems can be constructed based on the ideas of Contrastive Hebbian Learning~\cite{movellan1991contrastive}. These approaches include Equilibrium Propagation~\cite{scellier2017equilibrium}, Coupled Learning~\cite{stern2021supervised} and Hamiltonian Echo Backpropagation~\cite{lopez2023self}.  All of these local learning rules are based on the comparison of two states of the system: (a) a \emph{free state} where only inputs are applied, and (b) a \emph{clamped state} where the outputs are nudged toward the desired values (Fig.~\ref{fig:Schematic}b). Below, we show how such learning rules change the physical system (Fig.~\ref{fig:Schematic}c), aligning its response eigenspace with the learned task, and increasing its responses to random forces (Fig.~\ref{fig:Schematic}d).
	
	More precisely, consider a noiseless physical Hamiltonian network with fixed learning degrees of freedom $\vec{w}$. The network dynamics and responses to external inputs are fully determined by the physical cost function $E(\vec{x};\vec{w})$ which controls how the  physical degrees of freedom $\vec{x}$ respond to external inputs.
	In the absence of such inputs the system equilibrates, locally minimizing the physical cost function to settle into a \emph{native state} $\vec{x}^0(\vec{w})$ in which forces (for mechanical networks) or currents (for flow or electrical networks) are balanced $\vec{\nabla}_{\vec{x}} E(\vec{x},\vec{w})|_{\vec{w}} = 0$, or, in components,  $\partial E/\partial x_a|_{\vec{w}}=0$ for all $a$.
	Below we will consider small perturbations around such native states; that is, we discuss learning in the limit of linear response, by assuming that the forces applied to the system as inputs are weak in the sense that they cause small physical deformations. The learning approaches we discuss are practical and effective well beyond linearity~\cite{stern2021supervised}. However, in the linear regime we will show that learning is dominated by characteristic network phenomenology including reduction of the effective dimension of responses, softening of the system, and alignment of dynamics with the learned task.  Beyond linear response, other mechanisms, such as multi-state learning~\cite{stern2020continual, rouviere2021emergence, falk2023learning}, also become relevant.
	
	In  linear response it suffices to consider an expansion of the physical cost function around the native state up to the first non-vanishing, i.e., second, order:
	\begin{equation}
		E(\vec{x},\vec{w}) \approx E(\vec{x}^0,\vec{w}) + \frac{1}{2} (\vec{x} - \vec{x}^0)^T H(\vec{w}) (\vec{x} - \vec{x}^0)
		\label{eq:2.1}
	\end{equation}
	where the superscript $T$ denotes the transpose, and $H$ is a (symmetric) \textit{physical Hessian} matrix, the components of which are $H_{ab}(\vec{w}) = \partial^2 E(\vec{x},\vec{w}) / \partial x_a \partial x_b |_{\vec{x} = \vec{x}^0}$. In the following we name this matrix the Hessian as a shorthand.
	The first order term in the Taylor expansion vanishes, since the native state $\vec{x}^0(\vec{w})$ is a minimum of the physical cost function. The Hessian is a function of $\vec{w}$ both explicitly, and implicitly through the dependence of $\vec{x}^0$ on the learning degrees of freedom $\vec{w}$.
	
	To simplify the language in the remainder of this paper, we will use ``force" to denote forces in the case of a mechanical network or currents in the case of a flow network. 
	
	\subsection{Training physical responses at network nodes}
	
	Consider a generalized external input force $\vec{F}$ applied to the physical degrees of freedom, namely tensile forces in a mechanical network, or a set of currents in a flow network, applied to specific nodes.  This force could be applied locally (at a subset of nodes) or globally, e.g., a ``compression'' applying forces to all the $\vec{x}$ toward a certain point. These input forces will affect the physical cost function, prompting the system to equilibrate in a new
	\emph{free state} $\vec{x}^F$
	which minimizes the free state physical cost function:
	\begin{equation}
		\begin{aligned}
			%E^F (x_a;w_i )&=E^0+\frac{1}{2} (x_a-x_a^0 ) H_{ab} (x_b-x_b^0 )-F^I_a x_a \\
			E^F (\vec{x},\vec{w})&=E(\vec{x},\vec{w})- %\vec{F}^I 
			\vec{F}
			\cdot \vec{x} \\
			\frac{\partial}{\partial x} E^F (\vec{x},\vec{w}) &= 0 \ \  \Longrightarrow  \ \ 
			\frac{\partial}{\partial x} E(\vec{x},\vec{w}) = 
			\vec{F}
		\end{aligned}
		\label{eq:2.a1}
	\end{equation}
	In the linearized approximation (\ref{eq:2.1}), this gives
	\begin{equation}
		\vec{x}^F - \vec{x}^0 = H^{-1} 
		%\vec{F}^I
		\vec{F}
		\label{eq:xf}
	\end{equation}
	where we used the fact that the Hessian matrix is symmetric.  In other words, the system responds by shifting the native state by the inverse Hessian applied to the input force. %by applying the inverse Hessian $H^{-1}$ around the native state to the applied input force $F^I$. 
	%Note that $\vec{x}^F$ depends on the learning degrees of freedom $\vec{w}$ explicitly through the Hessian, and implicitly through the native state $\vec{x}^0(\vec{w})$.
	
	Thus, deformations around the native state with small Hessian eigenvalues are ``soft'' -- they exhibit a larger response to applied forces. Note also that the entries of the inverse Hessian depend globally on all the learning parameters. So the change produced by an external force on a given physical degree of freedom  can depend non-locally on the values of all the learning degrees of freedom, and not just on, say, the weights of edges connected to the network node in question.
	
	In order to proceed, we must define a task in terms of desired outputs in response to the inputs, which we can express in terms of physical constraints that must be satisfied. The constraints can be local, applying to a subset of nodes designated as output nodes, or global, applying to all of the nodes. Because the response is expressed in terms of the physical degrees of freedom, the constraints can be defined by demanding that the free state response $\vec{x}^F-\vec{x}^0$ satisfies a desired relationship encoded in a functional $c(\vec{x}^F-\vec{x}^0)=0$. For example, if one desires $\vec{x}^F-\vec{x}^0$ to equal $B$ at a given output node $o$, then the functional is simply $x^F_o-x^0_o-B$. This can be thought of as a single basic task. More complicated tasks can be defined by adding functionals for many such tasks. For example, a classification task may be posed such that all members of each class (each set of inputs in the class) prompt a response that satisfies a particular constraint for that class.  
	
	%To define a learning task, consider an output constraint which the system should implement in response to a given input force,  i.e. some function $f(\vec{x})=0$ defining a relationship between the physical degrees of freedom. The constraint may be local or global, i.e., constraining each degree of freedom $x_a$ separately or in combination. We can define tasks by demanding that the free state responses $\vec{x}^F-\vec{x}^0$ satisfy the target constraints $f(\vec{x}^F-\vec{x}^0)=0$.  More complicated tasks can be defined by adding more target input-output relationships. For example, we can define a classification task by requiring that each input force in a particular class should lead to responses  satisfying the same output constraint.
	
	Suppose we have multiple tasks, i.e., pairs of input forces and output constraints:  $(\vec{F}_{r}, c^{(r)}(\vec{x}^F_{r}-\vec{x}^0))$ indexed by $r = 1 \cdots n_T$. 
	Since we are studying the linear response regime, we can linearize any constraint in terms of the free state response as $c^{(r)} \approx \vec{A}_{r}^T(\vec{x}^F_r-\vec{x}^0) - B_r$, with a vector $\vec{A}_{r}$ and a scalar $B_r\geq 0$ determining the desired response. The weak input force regime dictates that, at least prior to training, the system responds much more weakly than desired, i.e. $\vert\vert\vec{A}_{r}^T(\vec{x}^F_r-\vec{x}^0)\vert\vert \ll B_r$. This is a sensible regime for untrained random systems, specifically over-constrained flow and elastic networks, which typically have a very weak coupling between putative input and output sectors, specifically if they are physically distant from each other.
	
	We can quantify how well the system performs the tasks, i.e., implements these response constraints,  in terms of a \textit{learning cost function} $C$.  For example, we can use a Mean Squared  Error (MSE) cost:
	\begin{equation}
		\begin{aligned}
			C(\{ \vec{x}_r \}, \vec{x}^0(\vec{w})) \equiv \frac{1}{2} n_T^{-1}\sum_r [c^{(r)}(\vec{x}_{r}-\vec{x}^0)]^2  
		\end{aligned}
		\label{eq:2.a2}
	\end{equation}
	evaluated at the free state responses $\vec{x}_r = \vec{x}^F_r$. For a random, untrained physical system, the task has nothing to do with the structure of the network, and so the free state responses will be diffusely and weakly spread over the network.   As a result free state does not move much from the native state, and the cost function will be dominated by $B_r$.  This gives $C\approx 0.5 n_T^{-1}\sum_r B_r^2$ which will tend to be high. Learning is the process of modifying the system, changing its learning degrees of freedom such that the cost function is reduced. How can this be done?
	
	Computational machine learning algorithms generally minimize the learning cost function by performing gradient descent on $C$. Typically, this means that global information is required to determine local changes in the network. By contrast, a physical learning system must use local learning rules to achieve the same goal.
	We take the approach of Contrastive Learning in its incarnation as Equilibrium Propagation~\cite{scellier2017equilibrium}. In this approach, the local learning rule effectively approximates gradient descent on the learning cost function.
	
	First consider training a system for a single task defined by the target constraint $c(\vec{x} -\vec{x}^0)=0$.  Suppose we apply the input force $\vec{F}$ and the system equilibrates at some free state $\vec{x}^F$. To train the system we could nudge it towards the desired output by applying an additional weak {\it output force} 
	\begin{equation}
		\eta \vec{F}^O = -\eta \, \frac{\partial}{\partial x} C |_{\vec{x} = \vec{x}^F} = \eta \, [\vec{A}^T (\vec{x}^F-\vec{x}^0) - B]\vec{A} ,
	\end{equation}
	where $\eta$ is a small parameter that we have explicitly separated out, and the last equation applies to a quadratic cost function (Eq.~\ref{eq:2.a2}) and linearized constraint.  The physical system responds to the nudge by settling into a {\it clamped state} $\vec{x}^C$ satisfying
	\begin{equation}
		\vec{x}^C - \vec{x}^F = \eta \, H^{-1} \vec{F}^O \, .
		\label{eq:2.a4}
	\end{equation}
	The clamped state depends on the learning degrees of freedom $\vec{w}$ explicitly through the Hessian and implicitly through the additional dependencies in the free state $\vec{x}^F$.
	We can describe this equivalently by saying that the  system minimizes a {\it clamped} physical cost function
	\begin{equation}
		E^C (\vec{x},\vec{w})=E^F(\vec{x},\vec{w})+\eta C(\vec{x}, \vec{x}^0(\vec{w})) %= E(\vec{x},\vec{w}) - \vec{F} \cdot \vec{x} + \eta  \, C
		\label{eq:2.a3}
	\end{equation}
	Thus the clamped state is the free state, nudged slightly by an extra output force  related to a learning cost function that arises if the system does not satisfy the desired constraints.

	The contrastive learning approach compares the free and clamped states to derive an approximation to the gradient of the learning cost function that can be minimized more readily via local learning rules. Define the contrastive function:
	\begin{equation}
		\begin{aligned}
			\mathcal{F}\equiv \eta^{-1}[E^C(\vec{x}^C,\vec{w}) - E^F(\vec{x}^F,\vec{w})]
		\end{aligned}
		\label{eq:2.a5}
	\end{equation}
	Previous work has showed that the partial derivative of the contrastive function with respect to the learning degrees of freedom $\vec{w}$ approximates the gradient of the learning cost function $C$ in the limit $\eta \to 0$~\cite{scellier2017equilibrium}:
	\begin{equation}
		\begin{aligned}
			\frac{dC}{d\vec{w}} = \lim_{\eta\to 0} \frac{\partial}{\partial w}\mathcal{F}
		\end{aligned}
		\label{eq:2.a6}
	\end{equation}
	On the right hand side we differentiate only the explicit $\vec{w}$ dependencies in the contrastive function, and not the implicit dependencies via the solutions for the free and clamped states.  We will also assume that the MSE cost function $C$ does not depend directly on $\vec{w}$, as it is a combination of physical constraints. The only explicit dependence of the physical cost function on the learning degrees of freedom then appears in the physical Hessian $H=H(\vec{w})$.
	
	Using (\ref{eq:2.a1}) and (\ref{eq:2.a3}) for $E^F$ and $E^C$ in $\mathcal{F}$, the Taylor expansion of the physical cost function in (\ref{eq:2.1}), and the linearized approximations for $x^F$ and $x^C$ in (\ref{eq:xf}) and (\ref{eq:2.a4}) gives
	
	\begin{equation}
		\begin{aligned}
			\frac{\partial}{\partial w}\mathcal{F} &= \eta^{-1} \frac{\partial}{\partial w}[E^C(\vec{x}^C;\vec{w}) - E^F(\vec{x}^F;\vec{w})] \\
			& \approx 
			\vec{F}^T \, H^{-1} (\frac{\partial}{\partial w} H)  H^{-1} \vec{F}^O  \approx \\
			&\approx B \vec{F}^T \, H^{-1} (\frac{\partial}{\partial w} H)  H^{-1} \vec{A}
			%\\
			%&\approx \frac{1}{2}[\vec{F}^O \otimes\vec{F}  + \vec{F} \otimes\vec{F}^O] H^{-1} (\nabla_{\vec{w}}H) H^{-1}
		\end{aligned}
		\label{eq:2.a7}
	\end{equation}
	where we kept only the term that survives the $\eta \to 0$ limit in the second line, and used the fact that $H$ and $H^{-1}$ are symmetric matrices. Note that the approximation in the second line, that the output force is approximately constant in each learning step, is valid as long as the free state response remains weak (and $C$ is relatively high). This will be true in the early stages of training because the network structure is not adapted to the task, but at the end of training this approximation fails because the network will respond strongly in the desired manner.  But the output force also approaches zero in this limit, and so learning concludes successfully.  In other words, the approximation above describes the critical stages of learning, and when it fails the network has already learned the task.

	Learning now proceeds by following this derivative of the contrastive function with a learning rate $\alpha$
	\begin{equation}
		\begin{aligned}
			\delta \vec{w} = -\alpha \frac{\partial}{\partial w}\mathcal{F}%= - \frac{\alpha}{2}[F^O_c F^I_d  + F^I_c F^O_d] H^{-1}_{ca} \frac{\partial H_{ab}}{\partial w_i} H^{-1}_{bd}
		\end{aligned}
		\label{eq:2.a8}.
	\end{equation}
	This learning rule has two key properties. First, learning is local. Every learning degree of freedom is modified according to the local difference between the free and clamped values of the physical cost, spatially localized at that edge of the network~\cite{scellier2017equilibrium, stern2021supervised}. Second, the rule for modifying $\vec{w}$ is proportional to the alignment of the input force $\vec{F}$ and the desired response $\vec{A}$ in an inner product determined by the symmetric matrix $H^{-1} \nabla_{\vec{w}} H H^{-1}$. We will see later how this property causes a realignment of the inherent physical coordinate system. 
	%Each learning degree of freedom tends to decrease if these vectors align (as $H$ is positive semi-definite), and increase otherwise. 
	
	\added{Next, we must choose a model for the physical Hessian and its dependence on the learning degrees of freedom. The general form of the Hessian we consider is 
		\begin{equation}
			\begin{aligned}
				H_{ab}&(\vec{w}) = \frac{1}{2}\sum_i [ L_{ai}\phi_i(w_i) R_{ib} + R^T_{ai}\phi_i(w_i)L^T_{ib} ] \equiv \sum_i h_{ab}^i\\
				\delta w_i &\approx -\frac{\alpha B \phi_i'}{2} \sum_{ab} (H^{-1}\vec{F})^T_a \, [ L_{ai} R_{ib} + R^T_{ai}L^T_{ib} ] (H^{-1} \vec{A})_b \, ,
			\end{aligned}
			\label{eq:2.Hess}
		\end{equation}
		where $L_{ai}, R_{ia}$ are left and right matrices selecting the physical degrees of freedom that participate in interaction $i$, and $\phi_i(w_i)$ a (possibly nonlinear) edge-wise function relating a `bare' learning degree of freedom $w_i$ with the physical Hessian. $\phi_i'$ is the partial derivative of the transformation $\partial_{w}\phi_i |_{w_i}$, particular to that interaction $i$. This form of the Hessian is inspired by linear physical system of interest like Ising models, as well as flow and mechanical networks, where it is possible for distinct linear combinations of physical degrees of freedom to generate interactions. The nonlinearity can, for example, represent a reparameterization of the conductance in a flow network in terms of resistance. The second line in Eq.~\ref{eq:2.Hess} is derived in detail in Appendix B.}
		
		\added{Note that the physical Hessian must be real-valued, symmetric and positive definite. The symmetry is guaranteed by definition and real values are also easily guaranteed, but one should be careful that the choices of matrices $L,R$ and the functions $\phi_i(w_i)$ result in positive definiteness. This form of the Hessian supports a general construction of linear physical models, with possible nonlinear transformation of the learning degrees of freedom. In particular, Eq.~\ref{eq:2.Hess} supports physical networks of interest, as we discuss next.}
	
\begin{figure}
		\includegraphics[width=0.95\linewidth]{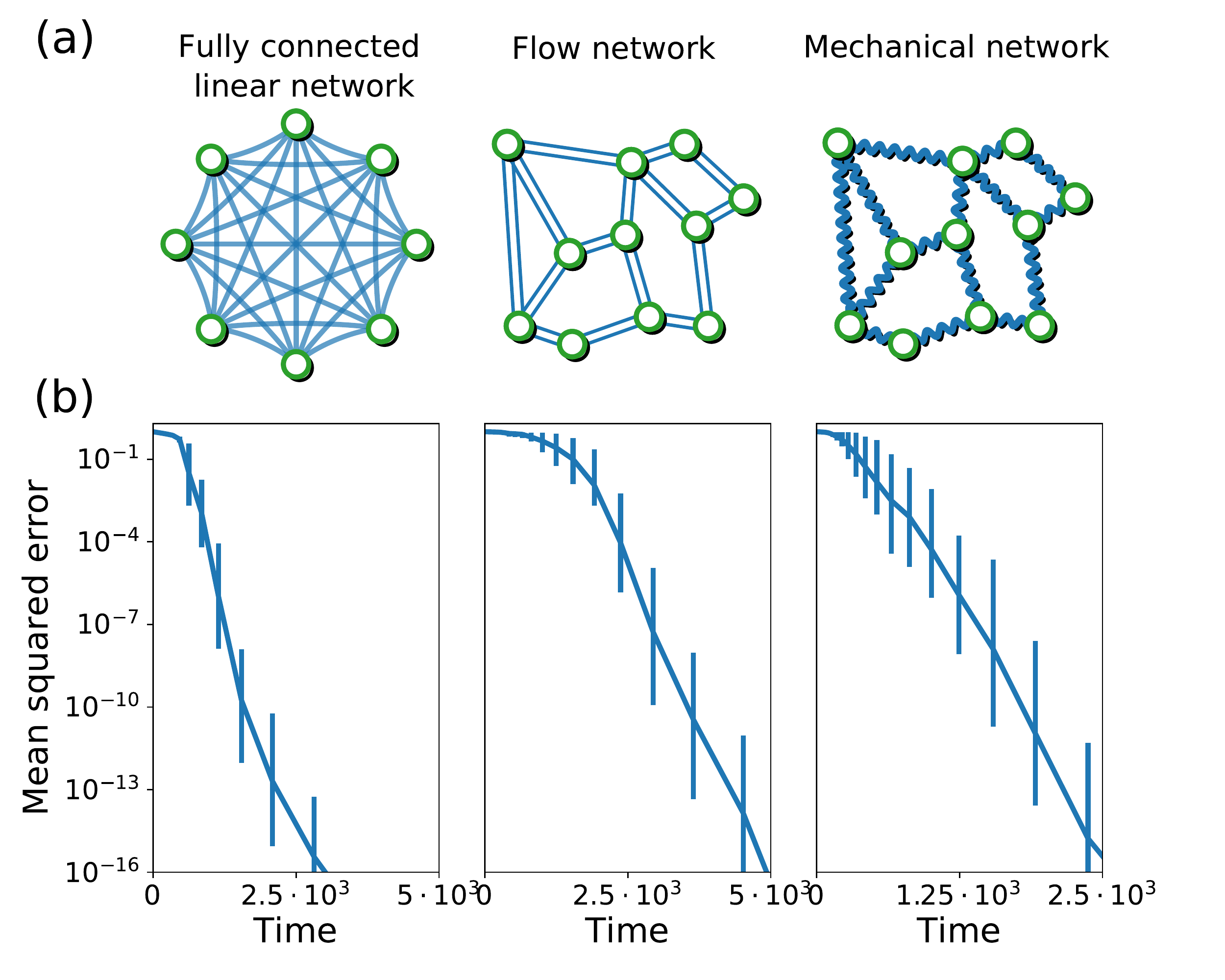}
		\caption{Training physical networks. a) We train several types of physical networks in the linear response regime, in particular fully connected linear networks (where physical degrees of freedom are node values), Flow networks and mechanical spring networks (where physical responses are naturally represented as difference over edges). b) Training these systems on single tasks, we generally find that local learning rules succeed, reducing the mean squared error by many orders of magnitude (geometric mean over 50 realizations of networks trained with single tasks).
			\label{fig:Errors}}
	\end{figure}

	To illustrate, consider a model where $\vec{x}$ are physical variables at $N$ nodes, while the $N_w = \frac{1}{2}N(N-1)$ learning parameters are at edges linking each pair of nodes.  In this fully connected case, the learning parameters are naturally represented as a matrix $W$ with entries $W_{ab}$, where $a,b = 1 \cdots N$ (Fig.~\ref{fig:Errors}a). 
	The energy (physical cost) function for such a {\it node network} is a function of the node variables: 
	\begin{equation}
		\begin{aligned}
			E(\vec{x};W) = \frac{1}{2}\vec{x} H(W) \vec{x}  .
		\end{aligned}
		\label{eq:2.a9}
	\end{equation}
	The Hessian of this energy is $H_{ab}(W_{ab}) = \frac{1}{2}\sum_{ij}[\delta_{ai}w_{ij}\delta_{jb} + \delta_{bi}w_{ij}\delta_{ja}]=\frac{1}{2}[W_{ab}+W_{ab}^T]$, and is thus linear in the learning degrees of freedom.
	%For sufficiently small forces compared to the scale set by the Hessian, any classical Hamiltonian system can be written in this way, and learning is the process of modifying the edge weights to amplify desired responses around rest, while suppressing unwanted responses to inputs.
	Learning is the process of modifying the edge weights to amplify desired responses around rest, while suppressing unwanted responses to inputs.
	For this model the learning rule for a given edge weight $W_{ab}$ is
	\begin{equation}
		\begin{aligned}
			%\delta{W_{ab}} &= - \frac{\alpha}{2} \vec{F}^T W^{-1}  \,
			%\partial_{W_{ab}}W \,  W^{-1} \vec{F}^O = \\
			%&=- \frac{\alpha}{2} (W^{-1}\vec{F})_a(W^{-1}\vec{F}^O)_b + {\rm transpose}   
			\delta{W_{ab}} \approx - \frac{\alpha B}{2} (W^{-1}\vec{F})_a(W^{-1}\vec{A})_b + {\rm transpose}   
		\end{aligned}
		\label{eq:2.a10}
	\end{equation}
	Here, we used the fact that the $a,b$ and $b,a$ entries of the matrix $\partial_{W_{ab}} W$ equal 1, while all the other entries are 0.  So in the second line we are multiplying the $a^{\rm th}$ component of the vector $W^{-1}\vec{F}$ by the $b^{\rm th}$ component of $W^{-1}\vec{A}$ and adding the transpose.

	To show that the physically realizable learning rule  Eq.~\ref{eq:2.a10} converges, we first tested it on a relatively easy task. We initialized networks with $N=20$ nodes with energy functions of the form (\ref{eq:2.a9}) and weights $W_{ij}$ drawn randomly from a standard Gaussian $\mathcal{N}(0,1)$. To ensure a positive definite Hessian, we explicitly symmetrized the weights and added a term proportional to the identity $\delta_{ab}$, $H_{ab} = \frac{1}{2}(W_{ab}+W_{ab}^T) + 3\sqrt{2N}\delta_{ab}$. %Then $H=W$ (since W is already symmetric) and all its eigenvalues are positive. 
	The resulting random Hessians initially have eigenvalues in the range $\sim (10,27)$. We then defined a random learning task by picking an input force  with each component drawn from a standard Gaussian, and a linear output constraint $c(\vec{x} -\vec{x}^0) = \vec{A}^T (\vec{x} - \vec{x}^0)  - B$ with the entries of $\vec{A}$ and $B$ also drawn from standard Gaussians (see Appendix A).  The local learning rule was effective in training such networks, reducing the error in Eq.~\ref{eq:2.a2} by many orders of magnitude (Fig.~\ref{fig:Errors}b).  Appendix A gives examples of other tasks such as allostery, regression and classification that can be trained in this way.

	%\subsection{Training mechanical and resistor networks}
	%\label{sec:trainingMechNetworks}
	
	In the mechanical/flow/electrical-resistor networks of interest to us, the physical cost function minimized by the network is defined via differences in the physical degrees of freedom $\vec{x}$ between nodes connected by edges whose non-negative learning degrees of freedom are $\vec{w}$. %In mechanical/flow networks, the energy/power is a sum over edges of the squared strains in springs (or voltage drops in resistors) weighted by the spring constant/conductance $\vec{w}$ across the edge.% Therefore, it is more appropriate to work in the space of differences between physical node variables connected by edges (Fig.~\ref{fig:Errors}a). Similar to networks discussed above, mechanical and resistor networks respond linearly around their native states.
	%In certain physical systems, the energy minimized by the network is a function of differences in the physical degrees of freedom $\vec{x}$ at nodes connected by edges, that in turn carry the learning degrees of freedom $\vec{w}$.  
	Each element $x_a$ of $\vec{x}$ can itself be an element of some d-dimensional vector space describing the physical attributes of nodes.  For example, we will consider mechanical networks where the node variables $x_a$ correspond to positions $(y^1_a, y^2_a,\cdots y^d_a)$ in a $d$-dimensional physical space, while the weights $w_i$ are spring constants (stiffnesses). We  use the notation that the product of two node variables $x_a x_b$ should be understood as an inner product in the physical space. For example, if the physical space is d-dimensional Euclidean space then $x_a x_b = \sum_{k=1}^d y^k_a y^k_b$. Likewise in resistor/flow networks, the $x_a$ describe node voltages/pressures $(V_a)$, which we will regard as one-dimensional vectors, and the $w_i$ describe  conductances on edges.  
	The energy (power) in these networks is a weighted sum over edges of the squared strains for springs, or squared voltage drops for resistors. In view of this, it is more  convenient to work with the {\it differences} between physical node variables connected by edges (Fig.~\ref{fig:Errors}a) rather than the node values themselves.

	To this end, we arbitrarily assign an orientation to every edge and define $\Delta_{ia}$ to take the values $\pm 1$ depending on whether the node $a$ is at the incoming or outgoing end of edge $i$, and $0$ otherwise.  The $N_w \times N$ quantities $\Delta_{ia}$ form the \textit{incidence matrix} $\Delta$, such that $\Delta \vec{x}$ is an $N_w$ dimensional vector, with entries that are the differences between the node variables on either side of each edge. In terms of this incidence matrix, we can write the physical cost function minimized by such {\it difference networks} as 
	\begin{equation}
		\begin{aligned}
			%E(\vec{x},\vec{w})\approx E(\vec{x}^0,\vec{w}) +\frac{1}{2} (\vec{x}-\vec{x}^0)^T \Delta^T \textrm{diag}(\vec{w}) \Delta (\vec{x}-\vec{x}^0)
			E(\vec{x},\vec{w})\approx \frac{1}{2} (\vec{x}-\vec{x}^0)^T \Delta^T \textrm{diag}(\vec{w}) \Delta (\vec{x}-\vec{x}^0)
		\end{aligned}
		\label{eq:2.b1}
	\end{equation}
	where $\textrm{diag}(\vec{w})$ is a diagonal matrix with entries $w_i$ which measure the conductances (for resistance/flow networks) or stiffness (for central-force spring networks) of the $N_w$ edges. Here, if the edge $i$ is incident on nodes $a$ and $b$, then the $i^{\rm th}$ component of $\vec{\delta} \equiv \Delta (\vec{x}-\vec{x}^0)$ is $(x_a - x^0_a) - (x_b - x^0_b)$.  In terms of these differences the second term in the physical cost function (\ref{eq:2.b1}) is $\sum_{i} w_i \delta_i \cdot \delta_i$.  Since the $\delta_i$ are differences of d-dimensional physical node variables, as described above, their product is defined by an inner product in the physical space. We can map the difference network (\ref{eq:2.b1}) to the general Hessian model (\ref{eq:2.Hess}) by the identification 
	\begin{equation}
		H = \Delta^T\textrm{diag}(\vec{w})\Delta  \, .
		%\longleftrightarrow W
		\label{eq:Widentification}
	\end{equation}

	In this case, $R=L^T=\Delta, \phi(w)=w$. 
	%We can thus identify the physical Hessian as $H_{ab}(\vec{w}) = \partial^2 E(\vec{x},\vec{w}) / \partial \vec{x}_a \partial \vec{x}_b |_{\vec{x} = \vec{x}^0}=\Delta^T_{ai} \textrm{diag}(w)_{ij}\Delta_{jb}$, with 
	$\partial H_{ab}/ \partial w_i=\Delta^T_{aj}(\delta^{ii})_{jk}\Delta_{kb} $ where $\delta^{ii}$ is a $N_w\times N_w$ matrix with a $1$ in the $i^{\rm th}$ diagonal entry, and zeros elsewhere.   We similarly define the learning task as the satisfaction of a linear constraint,  $c(\vec{x}^F-\vec{x}^0)=0$.  In terms of these constraints we define a learning cost function $C=0.5c^2$.  Similarly to Eq.~\ref{eq:xf}, given an input $\vec{F}$, the free state response is $\vec{x}^F-\vec{x}^0 = H^{-1} \vec{F}$. The clamped state response is also the same as before, $\vec{x}^C-\vec{x}^F= \eta H^{-1} \vec{F}^O$,
	%\begin{equation}
	%\begin{aligned}
	%\vec{x}^F-\vec{x}^0 = H^{-1} \vec{F}
	%\end{aligned}
	%\label{eq:2.b2}.
	%\end{equation}
	%We can think of this as an effective inverse Hessian $H^{-1}_\Delta \equiv (\Delta^T \Delta) H^{-1} (\Delta^T \Delta)$ acting on an effective force $\Delta \vec{F}$.
	%Now we can again compute the clamped state  as before
	%\begin{equation}
	%\begin{aligned}
	%\Delta (\vec{x}^C-\vec{x}^F) = \eta (H^{-1}_\Delta) \Delta \vec{F}^O
	%\end{aligned}
	%  \label{eq:2.b3},
	%\end{equation}
	with an output effective force $\vec{F}^O \equiv -\frac{\partial}{\partial x}C$. In mechanical spring networks, such forces cause the displacement of network nodes, while in flow or resistor nets, these forces may be understood as injecting currents to nodes.
	Using these definitions and Eq.~\ref{eq:2.a8}, we find that the learning degrees of freedom $\vec{w}$ change in a training step by
	\begin{equation}
		\begin{aligned}
			\delta w_i &\approx -\alpha B (\Delta H^{-1} \vec{F})_i (\Delta H^{-1} \vec{A})_i
		\end{aligned}
		\label{eq:2.b4}
	\end{equation}
	where on the right side we are simply multiplying the $i^{\rm th}$ component of the vectors 
	$\Delta H^{-1} \vec{F}$ and $\Delta H^{-1} \vec{A}$. Similar explicit learning rules where derived recently by Anisetti et. al.~\cite{anisetti2023learning}. We also see that  $\delta w_i$ in (\ref{eq:2.b4}) can be written as minus the product of the $i^{\rm th}$ components of the free and clamped state displacement difference at the $i^{\rm th}$ edge.  This means that the conductance/stiffness elements $\vec{w}$ tend to decrease if the response to the input and output forces at an edge align and increase otherwise.
	%Overall, learning in such physical networks where the relevant degrees of freedom are differences of node variables is similar to the networks discussed above in which the physical variables are the node variables themselves.  In particular, (\ref{eq:2.b4}) could be rendered in matrix form like (\ref{eq:2.a10}) if desired, except that the entries of the $W$ matrix in this case would be linear combinations of the network edge weights dictated by the incidence matrix. %The main change is that we must work in the space of differences $\Delta \vec{x}$, and hence with an  an effective inverse Hessian $H^{-1}_\Delta$.   

	In Fig.~\ref{fig:Errors}b, we show that Eq.~\ref{eq:2.b4} successfully trains $N=40$ flow networks and 
	%central force
	elastic 
	spring networks for allosteric tasks (see appendix A), reducing their error by multiple orders of magnitude. These networks were derived from Erdős–Rényi graphs with mean coordination number $Z=3$ for flow networks and $Z=4$ for $2$-dimensional mechanical networks.
	
	%In summary, in the remainder of the paper we will focus on these three example types of networks. (1) A fully-connected network whose physical cost function has the form of Eq.~\ref{eq:2.a9}, where $x_a$ is the physical degree of freedom for node $a$ and the learning degrees of freedom $w_i$ are organized in the Hessian $H_{ab}=W_{ab}$. These learning degrees of freedom evolve by Eq.~\ref{eq:2.a10}. (2) A flow or resistor network where the physical degrees of freedom are the node pressures or voltages and the learning degrees are the edge conductances. The physical cost function is described by Eq.~\ref{eq:2.b1} and the learning rule is described by Eq.~\ref{eq:2.b4}. (3) A mechanical central-force spring network, where the physical degrees of freedom $x_a$ are the node positions and the learning degrees of freedom are the spring constants of the edges. Again, the physical cost function is Eq.~\ref{eq:2.b1} and the learning rule is Eq.~\ref{eq:2.b4}. 

	\section{Trained networks are physically modified by learning}
	
	Above, we demonstrated a method for training physical networks to produce arbitrary linear responses to small inputs. We will show next that training modifies the physical properties of the network, changing the effective conductances, the dimension of the physical responses, and the alignment of the inherent coordinates of the physical response and the learned behaviors. 
	%In the following we discuss how learning modifies physical properties of the three example network systems described at the end of the last section.

	%As shown above, it is practically possible to train physical networks for arbitrary functions in the linear response regime (where input forces are small), allowing them to modify their own learning degrees of freedom to minimize the cost function. In the following we discuss how learning in such systems modifies their physical properties.
	
	%In linear response, we have seen that the physics of the network is fully determined by the Hessian at the native state. We therefore start this section by describing how the Hessian and its eigenspace change in response to learning. The physical properties of such systems that undergo learning can then be studied in relation to perturbations that are unrelated to the learned tasks. We study the dynamics of physical properties of interest due to learning, particularly the effective conductance of the system, the dimension of its physical response, and how the inherent coordinates of the system align to represent the learned behavior.
	
	\subsection{Changes to the Hessian and its eigenspace}
	
	Suppose a physical network implements the contrastive learning dynamics discussed above. Then, its changing responses to external forces are all manifested in modifications to the Hessian at the native state and its eigenspace.  The Hessian changes for two reasons: (a) it is a function of the network weights $\vec{w}$ which are changing, and (b) it is evaluated at the native state $\vec{x}^0$ which minimizes the physical cost function of the network, and hence is implicitly a function of the changing weights $\vec{w}$.
	%and ask how the system changes during learning. Since in linear response all physical effects are explained by the Hessian, let us describe how it changes in response to a learning step. 
	To evaluate these changes we first define the physical Hessian at a general network state $\vec{x}$  as $\tilde{H}(\vec{w}, \vec{x}) = \partial^2 E(\vec{x},\vec{w}) / \partial x_a \partial x_b |_{\vec{w}}$. Evaluating $\tilde{H}$ at the native state $\vec{x} = \vec{x}^0(\vec{w})$, which minimizes the physical cost function, gives the Hessian $H(\vec{w})$ in (\ref{eq:2.1}).
	
	In terms of $\tilde{H}$ the change in the Hessian can now be written as
	\begin{equation}
		\begin{aligned}
			\delta H = \left.
			\delta \vec{w} \cdot
			\frac{d\tilde{H}}{d\vec{w}} \right|_{\vec{x} = \vec{x}^0} 
			=
			\sum_i \delta w_i \,
			\left[\frac{\partial \tilde{H}}{\partial w_i}
			+ 
			\sum_a
			\frac{\partial \tilde{H}}{\partial x_a} 
			\frac{\partial x^0_a}{\partial w_i}
			\right]_{\vec{x} = \vec{x}^0}  
			%[\nabla_{\vec{w}}H+ \nabla_{\vec{x}}H\frac{d\vec{x}}{d \vec{w}}(\vec{x}^0)]\delta \vec{w}
		\end{aligned}
		\label{eq:3a1.1}
	\end{equation}
	%\begin{equation}
	%\begin{aligned}
	%\delta H = \frac{dH}{d\vec{w}}\delta \vec{w}=[\nabla_{\vec{w}}H+ \nabla_{\vec{x}}H\frac{d\vec{x}}{d \vec{w}}(\vec{x}^0)]\delta \vec{w}
	%\end{aligned}
	%  \label{eq:3a1}
	%\end{equation}
	In the right hand expression the sum on $i$ in the second term gives $\nabla_{w} x_a^0 \cdot \delta \vec{w}$, which is the change in the native state variable $x_a^0$ driven by learning.  %The expression $(\partial \tilde{H} / \partial x_a) (\partial x_a^0/\partial w_i)$ contains an implicit inner product in the physical space if the node variables $x_a$ are regarded as d-dimensional vectors (see Sec.~\ref{sec:trainingMechNetworks}.)
	
	For the linear systems of interest, with Hessians given by Eq.~\ref{eq:2.Hess}, $\tilde{H}$ is by construction independent of $\vec{x}$, and $\vec{x}^0$ is independent of $\vec{w}$. In these cases the second term in (\ref{eq:3a1}) vanishes and can be dropped. Such physical systems are in fact common: electrical resistor networks, as in Ref.~\cite{dillavou2022demonstration} always remain in the linear regime. Flow networks remain linear in the low Reynolds number regime, while initially unstrained spring networks are also linear over a reasonable range of strain when learning is performed on spring stiffnesses. In such systems the native state $\vec{x}^0$ does not depend on the learning degrees of freedom and stays fixed throughout the learning process. With this approximation we can compute the change in each component of the physical Hessian (see Appendix B for details).
	
	\begin{equation}
		\begin{aligned}
			\delta H_{ab} \approx -\alpha B\sum_{i,cd} [\frac{\partial h^i_{ab}}{\partial w_i}] [v \frac{\partial h^i}{\partial w_i} v^T]_{cd}  \frac{f_c}{\lambda_c} \frac{a_d}{\lambda_d} ,
		\end{aligned}
		\label{eq:3a1}
	\end{equation}
	
	where $H_{ab}=\sum_c v^T_{ac} \lambda_c v_{cb} $ is a diagonalization of the Hessian, with $\lambda_a, v_{ab}$ the eigenvalues and eigenmodes of the physical Hessian $H_{ab}$. Here $f_a = \sum_b v_{ab} F_b, a_a = \sum_b v_{ab} A_b$ are projections of the input and output vectors on each of the eigenmodes, and $\frac{f_c}{\lambda_c}, \frac{a_d}{\lambda_d}$ are these same projections, where each component is scaled by the associated inverse eigenvalue. 
	%Also note that all of these and the following expression are proportional to the learning rate $\alpha$.%, which for the sake of convenience we set $\alpha=1$, effectively defining the units of time.

	We see that regardless of the choice of model for the physical Hessian $H_{ab}$, the change in the Hessian due to learning is given by a sum over symmetric matrices, which yields a symmetric modification, as required. However, note that this learning rule does not guarantee that the Hessian remains positive definite.  Violation of positive semi-definiteness in this learning rule is a signature of the failure of the linear approximation.  For example,  if the learning  rule tries to push a parameter into an unphysical regime, the change simply will not happen. We are focusing on learning in the linear regime where these sorts of processes do not happen.

	We wrote above the modification of the Hessian given a single task, i.e., input force - output constraint pair. Each additional task that the network is trained for contributes such a modification, and therefore the total change in the Hessian due to a learning step consisting of $r = 1 \cdots n_T$  tasks is
	\begin{equation}
		\begin{aligned}
			\delta H_{ab} &= \sum_r \delta H_{ab}^{(r)} %= \\
			%&= -\frac{\alpha}{2}  \sum_r(H^{-1}\vec{F}_r)_a (H^{-1}\vec{F}^O_r)_b +\rm{transpose} \, .
			%&= -\frac{\alpha}{2 n_T} \Xi_{abcd}  \{H^{-1} \sum_r [F^O \otimes F^I + F^I \otimes F^O]_r H^{-1}\}_{cd}
		\end{aligned}
		\label{eq:3a3}
	\end{equation}

	\added{Now that we know how learning modifies the physical Hessian, we can track its evolution and predict the important features of the system in the neighborhood of the native state. To do so, we first discuss how the eigenspace of the Hessian changes in response to learning. Using first order perturbation theory, we can compute the changes in the eigenvalues $\lambda_n$ and eigenmodes $\vec{v}_{n}$ of the Hessian ($n = 1 \cdots N$) due to each task. To do this, we need to compute sums of the form 
		\begin{equation}
			\begin{aligned}
				\mathcal{M}_{mn}^{ab} \equiv \sum_{i,cdef} v_{ac} [\frac{\partial h^i_{cd}}{\partial w_i}] v_{db}^T \cdot v_{me} [\frac{\partial h^i_{ef}}{\partial w_i}] v_{fn}^T
			\end{aligned}
			\label{eq:MSums}
	\end{equation}}
	\added{In Appendix B we show how our choice of the Hessian model (Eq.~\ref{eq:2.Hess}), the fact that there are many interactions, coupled with some assumptions of their form, results in simplification of the sum in Eq.~\ref{eq:MSums}. We can approximate the sums $\mathcal{M}_{mn}^{ab}$, such that are only finite when indices are chosen identical in pairs, i.e. $\mathcal{M}_{mn}^{mn} \equiv X_{mn} > 0$, $\mathcal{M}_{mm}^{nn} \equiv Y_{mn} \geq 0$, which are all non-negative factors. With this in mind, we can estimate the change in the Hessian eigenvalues and eigenvectors using first order perturbation theory: }% Since the estimate in effect averages over many interactions at each node in the sums (\ref{eq:MSums}) we will use the angle-bracker notation $\langle \cdot \rangle$ for quantities computed in this way:
	
	\begin{eqnarray}
		\delta  \lambda_n &=& \vec{v}_n^T \delta H \vec{v}_n \nonumber \\
		&\approx &- \alpha B \sum_{m} Y_{nm} \frac{f_m a_m}{\lambda_m^2}
		\label{eq:3a6} \\
		\delta \lambda_{n}^{-1}  &=& -\lambda_n^{-2} \delta  \lambda_n \approx
		\nonumber \\
		&\approx& \frac{\alpha B}{\lambda_n^2} \sum_{m} Y_{nm} \frac{f_m a_m}{\lambda_m^2}\\
		\delta \vec{v}_{n}  &=& \sum_{m\ne n} 
		\frac{\vec{v}_m^T \, \delta H \, \vec{v}_n}{\lambda_n - \lambda_m} \vec{v}_{m}
		\label{eq:3a6.5}
		\nonumber \\
		&\approx& \alpha B \sum_{m\ne n}\frac{X_{mn} (f_m a_n + f_n a_m)}{\lambda_m\lambda_n(\lambda_m - \lambda_n)} \vec{v}_{m}
		\label{eq:3a7}
	\end{eqnarray}
	
	Note that the second order correction to the eigenvalues is $\delta \lambda_n^{(2)}\sim\alpha^2 \sum_{m\ne n} (\lambda_n - \lambda_m)^{-1}$~\cite{potters2020first}, and is thus negligible compared to the first order term ($\sim \alpha$) in the limit of slow learning rate $\alpha \ll 1$, except when two eigenvalues nearly cross.  At that point an effective ``repulsion'' prevents eigenvalue crossing. 
	
	%where the outer product $\otimes$ means that $\vec{F}_r \otimes \vec{F}_r^O$ is a matrix with components $F_{ra} F^O_{rb}$, and we assume a generic starting network with non-degenerate eigenvalues. If the network is fine tuned to have some identical eigenvalues we will have to employ the techniques of degenerate perturbation theory, at least in the first learning step.    The $\lambda^2$ and $\lambda_n\lambda_m$ in the denominators of (\ref{eq:3a6}) and  (\ref{eq:3a7}) arise from the action of the $H^{-1}$ factors in $\delta H$ (see Eq.~\ref{eq:3a3}) on the eigenvectors $\vec{v}_n$ and $\vec{v}_m$ in the numerators.
	
	%at which point a repulsive interaction prevents them from crossing.
	%. At the limit of slow learning rate $\alpha \ll 1$, it is negligible compared to the first order term ($\sim \alpha$), except when two eigenvalues nearly cross, at which point a repulsive interaction prevents them from crossing.
	
	\added{Two key points about the inverse eigenvalue corrections are: (1) Changes occur predominantly in the smaller eigenvalues because of the $\lambda_n^{-2}$ scaling. (2) We may define the \textit{alignment} of an eigenmode with the task by taking the product of its projections on the input and desired output vectors $\rho_n\equiv f_n a_n$. Since all of the factors $Y_{nm}$ are non-negative, if eigenmodes are positively aligned, specifically at the lower eigenvalues, then the eigenvalues are pushed down by learning (or inverse eigenvalues pushed up). The lower an eigenvalue is, the more important it is in determining the direction of these eigenvalue dynamics.}
	
	\added{The eigenvector dynamics in Eq.~\ref{eq:3a7} describe the rotation of the coordinate system of the physical Hessian. Each eigenvector changes by incorporating corrections proportional to all other eigenvectors, with contributions scaling based on their eigenvalues and projections on the input and desired output vectors. Note that these dynamics keep all eigenvectors normalized and orthogonal to each other. Eigenvectors associated with low eigenvalues tend to change more, and also more strongly affect the change in other eigenvectors, because of the dependence on the inverse eigenvalues $\lambda_m^{-1},\lambda_n^{-1}$. Also, we can show that these eigenvector dynamics imply the alignment of lower eigenvectors with the learned task. To see this, we compute the average change in the alignment of mode $n$ due to learning (see Appendix B for details):}
	
\begin{equation}
			\begin{aligned}
				\delta \rho_n \approx \alpha B \sum_{m\ne n}\frac{X_{mn} (f_m a_n + f_n a_m)^2}{\lambda_m\lambda_n(\lambda_m - \lambda_n)}
			\end{aligned}
			\label{eq:GenAlign}
	\end{equation}
	
	\added{In this expression, the sign of every member of the sum depends only on the sign of the difference in eigenvalues $\lambda_m - \lambda_n$, as all other expression are positive. This means, e.g. that the alignment of the lowest eigenmode $\rho_1$ is becomes more positive - the lowest mode aligns with the task. Recalling Eqs.~\ref{eq:3a6}-\ref{eq:3a7}, this stronger alignment will result in a more dominant effect of this mode on the eigensystem dynamics. In contrast, the modes associated with top eigenvalues are expected to misalign with the learned task, making them overall less important for the learning dynamics. It is notable that as these dynamics are symmetric with respect to the input force $f_n$ and the desired output vector $a_n$. Since these vectors are random with respect to the initial eigenvectors, we expect that the alignment of both of them with the lowest eigenmode,  will increase during learning, i.e. $\langle \delta f_1^2 \rangle \sim \langle \delta a_1^2 \rangle > 0$.  In another words, the lowest mode tends to rotate into the plane spanned by the input force and the output vector.  We also note that a fully trained system no longer changes its eigenvalues or eigenvectors because, while the input force $\vec{F}$ remains constant over training, the output force $\vec{F}^O=-\nabla_{\vec{x}} C$ diminishes and vanishes together with the learning cost function. Therefore, the approximations above are valid during most of the learning process, but not close to its conclusion.}
		
\added{Above we discussed the eigensystem dynamics for quite general Hessian models satisfying Eq.~\ref{eq:2.Hess}. We can use these results for the special cases of networks of interest to us. For the fully connected linear network given by Eq.~\ref{eq:2.a9}, the dynamics are considerably simplified and we can write closed form expressions:}
		
\begin{equation}
			\begin{aligned}
				\delta \lambda_n^{-1} &\approx \alpha B \frac{f_n a_n}{\lambda_n^4} = \alpha B \frac{\rho_n}{\lambda_n^4} \\
				\delta \vec{v}_n &\approx \alpha B \sum_{m\ne n}\frac{f_m a_n + f_n a_m}{\lambda_m\lambda_n(\lambda_m - \lambda_n)}\delta \vec{v}_m \\
				\delta \rho_n &\approx \alpha B \sum_{m\ne n}\frac{(f_m a_n + f_n a_m)^2}{\lambda_m\lambda_n(\lambda_m - \lambda_n)}
			\end{aligned}
			\label{eq:NodeDyn}
	\end{equation}
	
\begin{figure}
		\includegraphics[width=0.95\linewidth]{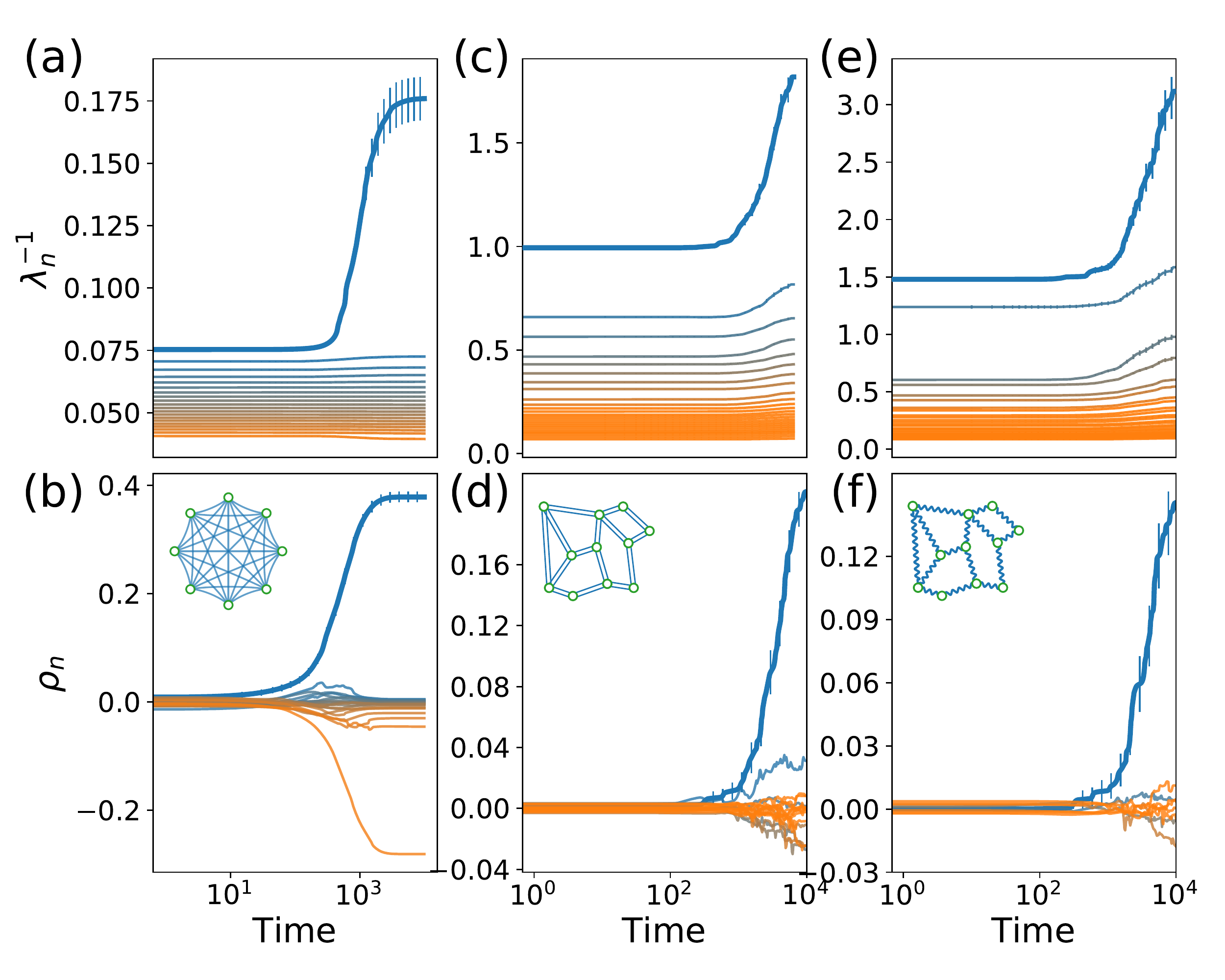}
		\caption{Hessian changes during learning. a) Hessian inverse eigenvalues change during training of fully connected linear networks ($N=20$) for a single task. The top inverse eigenvalue tends to significantly increase, showing that learning creates a soft mode. b) The lowest (blue) eigenmode of the Hessian significantly aligns with the input force force and output vector defined by the task. (c-f) Similar results are found for flow and mechanical networks ($N=40$), except that the higher eigenmodes do not align with the task. Results averaged over $150$ realization of networks and tasks.
			\label{fig:Hessian}}
	\end{figure}

	\added{These expressions clearly show the dynamics we described - aligned eigenmodes have their eigenvalues reduced (inverse eigenvalues increased), and low eigenmodes become more aligned with the task. Meanwhile,  eigenmodes with higher eigenvalues decrease their alignment over training, and we expect their associated eigenvalues to be further increased by learning. These considerations are verified in Fig.~\ref{fig:Hessian}a, where we simulate the Hessian dynamics of fully connected linear networks with physical cost function  (\ref{eq:2.a9}) during learning of one task. We see that the top inverse eigenvalue (bottom eigenvalue) is significantly increased, while the bottom inverse eigenvalue decreases in networks, here with $N=20$ nodes, averaged over $150$ realizations of networks and tasks. Moreover, by calculating the alignment of the eigenmodes $\rho_n$, we find that the bottom eigenmode aligns with the task and top eigenmodes mis-align with it (Fig.~\ref{fig:Hessian}b). Note that while we plot results averaged over realizations, the described scenario plays out in each individual learning simulation.}
	
	\added{For random difference networks (e.g., flow and elastic nets) of Eq.~\ref{eq:Widentification}, it is easiest to report the average eigenvalue dynamics as the geometry of interaction between nodes is more complicated than in the previous case. However, the results are still simplified because the left and right matrices are identical:
		\begin{equation}
			\begin{aligned}
				\delta \lambda_n^{-1} &\approx \frac{4 \alpha B}{\lambda_n^2} \frac{N_e}{N^2} [\frac{2f_n a_n}{\lambda_n^2} + \sum_{m\ne n} \frac{f_m a_m}{\lambda_m^2}]\\
				\delta \vec{v}_n &\approx 4 \alpha B  \frac{N_e}{N^2} \sum_{m\ne n}\frac{f_m a_n + f_n a_m}{\lambda_m\lambda_n(\lambda_m - \lambda_n)}\delta \vec{v}_m \\
				\delta \rho_n  &\approx 4\alpha B \frac{N_e}{N^2}\sum_{m\ne n}\frac{(f_m a_n + f_n a_m)^2}{\lambda_m\lambda_n(\lambda_m - \lambda_n)}
			\end{aligned}
			\label{eq:DiffDyn}
		\end{equation}
		where $N_e$ is the total number of edges (resistors or springs) in the network. The eigenvector and alignment dynamics behave similarly to the previous case, with the alignment of the lower modes increasing during learning. The eigenvalue dynamics are slightly different; a second term appears in the eigenvalue dynamics which is shared for all eigenvalues. This term tends to decrease all eigenvalues by the same amount if the eigenvectors (specifically the lower ones) tend to align with the task, and increase all of them otherwise. We therefore expect that in this case, contrary to the previous one, the alignment of the lower eigenvectors will cause  \textit{all} the eigenvalues to be decreased during learning by similar amounts. However, as in Eq.~\ref{eq:3a6.5}, the change in inverse eigenvalues is most pronounced in for the lowest modes that align with the task.}

	We verify these considerations for flow and mechanical networks with $N=40$ nodes in Fig.~\ref{fig:Hessian}c-f. The bottom eigenvalue is effectively reduced (inverse eigenvalue increased) by learning and the associated eigenmode aligns significantly with the input force (results averaged over $150$ realizations).

	\subsection{Effective conductance and dimension}
	
	The properties of a physical system are often characterized by its responses to generic forces (e.g., finite temperature fluctuations), quantified below by an {\it effective inverse-stiffness/conductance} $\bar{g}$.
	%(in the following we will call it the effective conductance).
	%A simple quantification of the response 
	%measure is the effective conductance $\bar{g}$, characterizing the response amplitude to typical forces. 
	Suppose we compute the responses to $M$ random forces $\{\vec{F}^R_m\}$ sampled from some distribution and indexed by $m$.  In each case we have the free state response $(\vec{x}^R-\vec{x}^0)_{m} = H^{-1}\vec{F}^R_{m}$. The effective conductance is the average amplitude of these responses:
	%is the amplitude of the physical response to random test forces averaged over the set of random forces:
	%. The response to such forces is computed similarly to what we have seen for the free state, given by $(\vec{x}^R-\vec{x}^0)_{m} = H^{-1}\vec{F}^R_{m}$. The effective conductance is the amplitude of the physical response to random test forces averaged over the set of random forces:
	\begin{equation}
		\begin{aligned}
			\bar{g} = M^{-1}\sum_m \frac{\vert \vec{x}^R_m-\vec{x}^0 \vert}{\vert \vec{F}^R_m \vert} %=  M^{-1}\sum_m \sqrt{\frac{\vert H^{-1}\vec{F}^R_m \vert^2}{\vert \vec{F}^R_m\vert^2}}
		\end{aligned}
		\label{eq:3b.1}
	\end{equation}
	
	%, $F^R_a\sim \mathcal{N}(0, N^{-1})$, where $N$ is the physical dimension of the system (number of physical degrees of freedom). 
	
	Suppose the random forces are drawn component-by-component independently from a Gaussian distribution and normalized to amplitude $\vert \vec{F}^R_m\vert^2=1$. Define the random force projections $\vec{f}^R\equiv  v \vec{F}^R$. As the eigenmodes are uncorrelated with the random forces, the components of $\vec{f}^R$ are inner products between random vectors on the unit sphere. In high dimension $N$, these are approximately drawn from a normal distribution $\mathcal{N}(0, N^{-1})$. The response of the system to the random input force is
	%Decompose the Hessian as $H^{-1}=v\Lambda^{-1}v^T$ ($v$ is a matrix of eigenmodes and $\Lambda$ a diagonal matrix of eigenvalues $\lambda$). The eigenmodes $v$ are a set of orthonormal vectors completely uncorrelated with the random forces.  Thus the components of the vector $R = v^T \vec{F}^R$ are inner products between random vectors on the unit sphere.  In high dimension $N$, these inner products are approximately drawn from a normal distribution $\mathcal{N}(0, N^{-1})$ with zero mean and variance $1/N$.  Now, note that
	
	\begin{equation}
		\begin{aligned}
			H^{-1}\vec{F}^R &= \sum_{b} v^T_{ab} \frac{f^R_b}{\lambda_b} \rightarrow \vert H^{-1}\vec{F}^R\vert^2  = \sum_a (\frac{f^R_a}{\lambda_a})^2
			%~~ \Longrightarrow = v\Lambda^{-1} R
			%\sim v\Lambda^{-1} \mathcal{N}(0, N^{-1})^N
			%\\
			%\vert H^{-1}\vec{F}^R\vert^2 & \sim R^T \Lambda^{-2} R 
			%(FH^{-1})(H^{-1}F)= F^R v \Lambda^{-1} v^T v \Lambda^{-1} v F^R
		\end{aligned}
		\label{eq:3b.2}
	\end{equation}
	where we used the fact that $v^T v$ is the identity.  The second line in (\ref{eq:3b.2}) equals the sum squared of the components of $\vec{f}^R$ scaled by the inverse square of the eigenvalues.  Thus its expectation value is a scaled sum of the variances of the components of $\vec{f}^R$, each of which equals $1/N$. We therefore find that the expected value of $\vert H^{-1}\vec{F}^R\vert^2$ is a sum over the square inverse eigenvalues and the effective conductance is simply
	\begin{equation}
		\bar{g} = \sqrt{\sum_a \lambda^{-2}_a}  \, .
		%\ \ \Longrightarrow \ \ 
		%\frac{\partial \bar{g}}{\partial \lambda_a} = -\frac{\lambda_a^{-3}}{\bar{g}}
		\label{eq:3b.3}
	\end{equation}
	As is well known, the effective conductance $\bar{g}$ is dominated by the lower eigenvalues. We saw that successful learning lowers the lowest eigenvalues, suggesting that trained systems have an increased effective conductance. Therefore, trained systems will be `softer' than random systems, exhibiting larger responses on average to random applied forces. Note that the increased effective conductance is unrelated to the specific details of the learned task; such physical systems trained for different types of tasks (regression, classification) are expected to become softer/more conductive (see Appendix A). Fig.~\ref{fig:PhysResponse}a shows how the effective conductance of the different physical networks studied rises during learning of a single task. These results are normalized so that the effective conductance at the beginning of learning is $\bar{g}(t=0)=1$, and averaged over $150$ different realizations of the network and training task. Here, the orange curves correspond to flow networks, the green ones correspond to mechanical networks, and the blue curves correspond to the fully-connected networks.
	
	\begin{figure}
		\includegraphics[width=0.95\linewidth]{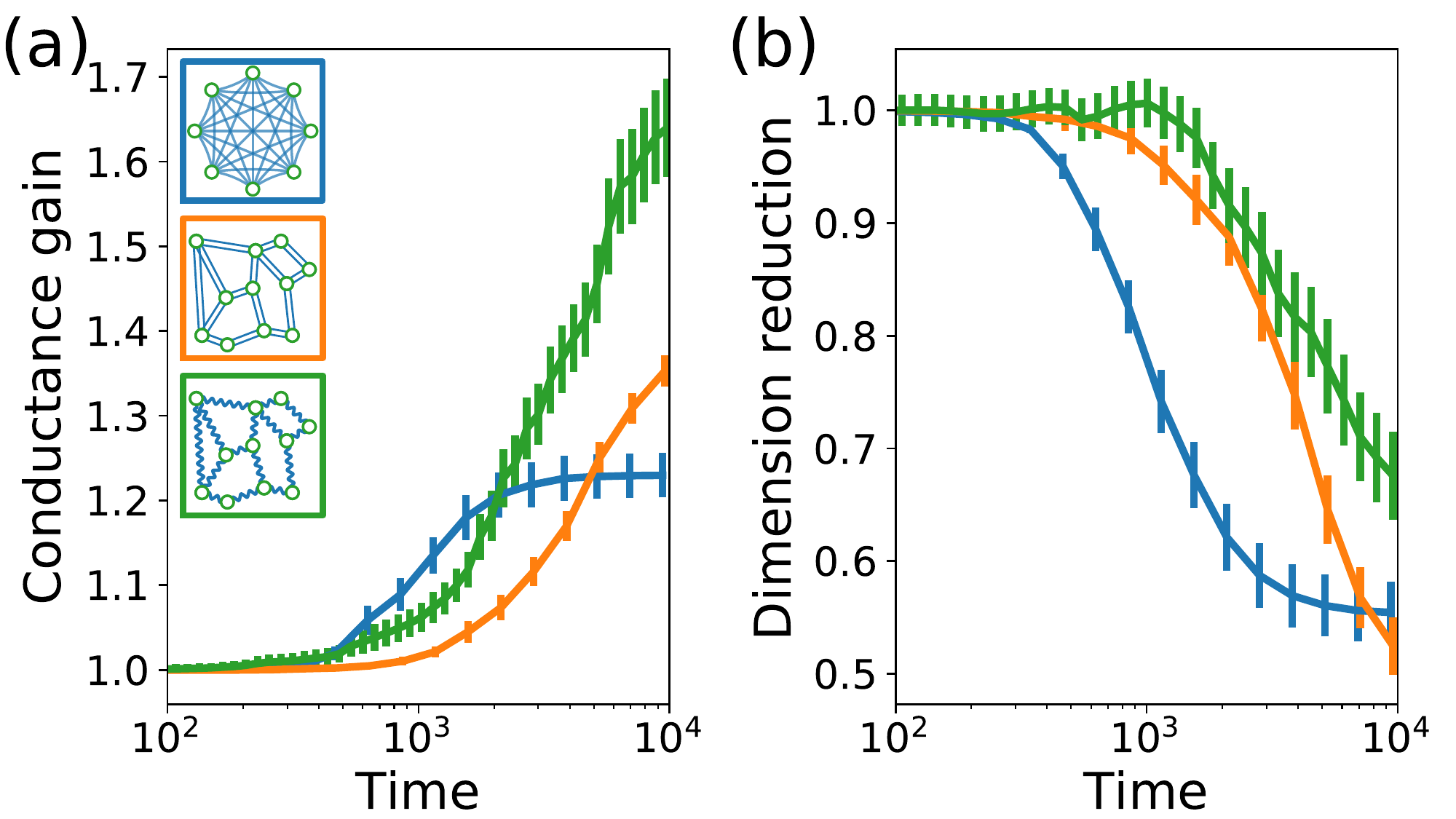}
		\caption{Training increases the effective conductance and reduces response dimension in physical networks. a) Effective conductance is increased during training in all systems considered, suggesting that trained systems are softer, with stronger responses to random forces. Lines show conductance values for different networks normalized to $1$ at initialization. b) The physical response dimension is decreased in all systems during learning, so that the response of these systems to random forces is concentrated in low-dimensional manifolds. Lines show physical dimension values normalized to $1$ at initialization. Results averaged over 150 networks and tasks. 
			\label{fig:PhysResponse}}
	\end{figure}

	%\subsection{Physical response dimension}
	
	We can also directly study the dimension of the space of responses to random forces.  While the system has $N$ physical degrees of freedom, typical responses are coupled, lowering the effective dimensionality, as has been observed in, e.g.,  proteins~\cite{leo2005analysis} and neural circuits~\cite{recanatesi2021predictive}. This effective dimension can be extracted by measuring how widely the responses are spread over the different eigenmodes of the Hessian. To define a measure of this spread, we again consider the response $\vec{x}^R_m-\vec{x}^0$ to random forces $\vec{F}^R_m$. Let $p_{am} \equiv \sum_{b}v_{ab}(\vec{x}^R_m-\vec{x}^0)_b $ be the projection of the responses onto the Hessian eigenvectors. Then the associated {\it participation ratio} is 
	\begin{equation}
		D_m \equiv \frac{\left[ \sum_a p_{am}^2 \right]^2}{\sum_a p_{am}^4}
		\label{eq:oneForceRatio}
	\end{equation}
	If a response $\vec{x}^R_m-\vec{x}^0$ is parallel to an eigenmode $\vec{v}_a$, it is orthogonal to the others, so $D_m = 1 $. On the other hand, if all eigenmodes participate in a given response with the same amplitude, i.e. $p_{am}=\pm\textrm{const}$, then $D_m = N$.  Thus $D_m$ captures the effective number of eigenmodes participating in the response to $\vec{F}^R_m$. Therefore it is natural to define the effective dimension of the response as
	\begin{equation}
		D_{{\rm eff}} =
		\frac{ \langle \sum_a p_{am}^2 \rangle^2}{\langle \sum_a p_{am}^4 \rangle}
	\end{equation}
	where the angle brackets denote an expectation value in the ensemble of random forces. In other words, we are defining the effective dimension as the ratio of the square of the second moment and the fourth moment of the projections of the response space onto the eigenmodes.  Note that this is similar to, but not the same as, another classic measure of response spread: the participation ratio (PR) dimension~\cite{xu2010anharmonic, recanatesi2022scale,sadjadi2021realizations}.  The latter measure simply takes the expectation value of (\ref{eq:oneForceRatio}) in the ensemble of random forces, rather than separately taking expectation values in the numerator and denominator.

	%The {\it participation ratio dimension} 
	%\begin{equation}
	%\bar{D} \equiv M^{-1} \sum_n D_m
	%  \label{eq:3c.1}.
	%\end{equation}
	%is the average of the effective response dimension $D_m$ over an ensemble of random forces.
	%as an effective way to count the number of participating eigenmodes, normalized as
	%\begin{equation}
	%\begin{aligned}
	%\bar{D} \equiv M^{-1}\sum_m \frac{[\sum_a (\vec{v}^T_{a} \cdot (\vec{x}^R_m-\vec{x}^0))^2]^2}{\sum_a (\vec{v}^T_{a} \cdot(\vec{x}^R_m-\vec{x}^0))^4}
	%\end{aligned}
	%  \label{eq:3c.1}.
	%\end{equation}
	%Here $m$ indexes a sum over responses to the $M$ random forces, and $a$ indexes the eigenvectors.
	%This PR dimension has reasonable limits. 
	%If a given response $\vec{x}^R_m-\vec{x}^0$ is parallel a given eigenmode $a$, then it is orthogonal to the others, and so  contributes of 1 to  $\bar{D}$.
	%we have $\vec{v}^T_{a} \cdot(\vec{x}^R_m-\vec{x}^0)=1$, then $D_m=1$.
	%This makes sense because the system response is parallel to a single eigenmode and orthogonal to all others.
	%On the other hand, if all eigenmodes participate in a given response with the same amplitude, i.e. $\vec{v}^T_{a} \cdot(\vec{x}^R_m-\vec{x}^0)=\pm\textrm{const}$ for every mode, then this response contributes $N$ to the $\bar{D}$.
	%and so the system has full dimensional response $D_m=N$. 
	
	Our effective dimension has a simple and intuitive expression in terms of the spectrum of eigenvalues of the Hessian (details in Appendix C):
	\begin{equation}
		D_{{\rm eff}} = \frac{(\sum_a \lambda_a^{-2})^2}{3 \sum_a \lambda_a^{-4}}
		\label{eq:3c.2}
	\end{equation}
	%\begin{equation}
	%\begin{aligned}
	%\bar{D}& = \frac{(\sum_a \lambda_a^{-2})^2}{3 \sum_a \lambda_a^{-4}}\\
	%\frac{\partial \bar{D}}{\partial \lambda_a} &\approx 4\lambda_a^{-5} \frac{\bar{D} - \lambda_a^2(\sum_a \lambda_a^{-2})}{\sum_a \lambda_a^{-4}}
	%\frac{\partial \bar{D}}{\partial \lambda_a} &\approx \frac{4D}{\lambda_a^5} \frac{\lambda_a^{-2} (\sum_a \lambda_a^{-2}) - (\sum_a \lambda_a^{-4})}{(\sum_a \lambda_a^{-2})(\sum_a \lambda_a^{-4})}
	%\end{aligned}
	%  \label{eq:3c.2}
	%\end{equation}
	If one eigenvalue is particularly small, it will dominate, and lead to an effective dimension of $1/3$. As we showed above, learning reduces the low eigenvalues of the physical Hessian, suggesting that learning reduces the physical response dimension for learned tasks. In fig.~\ref{fig:PhysResponse}b, we compute this effective dimension during learning for different physical systems (averaged over $150$ networks and tasks). We find that the physical dimension  decreases during learning as the system adapts to accommodate the learned task. There are other ways to estimate the dimension of the physical response -- in Appendix C we discuss some of them and show that for alternative definitions, the physical dimension is still reduced by learning.
	
	%To test how this changes the effective dimension we can take a derivative of (\ref{eq:3c.2}):
	%\begin{equation}
	%\frac{\partial D_{{\rm eff}}}{\partial \lambda_a} = \frac{4}{3\lambda_a^5} \frac{D_{{\rm eff}} - \lambda_a^2 \bar{g}^2}{\sum_a \lambda_a^{-4}}
	%\label{eq:DeffDeriv}
	%\end{equation}
	%We see that changing a large eigenvalue does not modify the effective dimension much because of the $\lambda_a^{-5}$ suppression.
	%The physical dimension change is strongly suppressed ($\sim\lambda_a^{-5}$) at higher eigenvalues, meaning that the change in the lowest lying eigenvalues dominates the change in the PR dimension. 
	%This makes sense because the system response to forces is controlled by the lowest lying eigenmodes -- indeed (\ref{eq:3c.2}) is dominated by the low eigenvalues. Conversely, decreasing the highest eigenvalues increases the effective dimension because this makes more eigenmodes active in the responses.
	%yet much smaller modification occurs when changing the highest eigenvalues. 
	%This shows that increasing (decreasing) the width of the Hessian eigenvalue spectrum would tend to decrease (increase) the physical dimension. 
	%In general, we saw that learning tends to decrease the lower eigenvalues of the Hessian eigenvalues, suggesting that learning reduces the physical response dimension for \emph{any} learned task.

	%\subsection{Physical alignment with the learned task}
	
	%As discussed above, the change in eigenmodes during learning tends to align the  modes corresponding to low eigenvalues with the manifold defined by the input and output forces. 
	
	\section{Training for multiple tasks}
	
	\begin{figure}
		\includegraphics[width=0.95\linewidth]{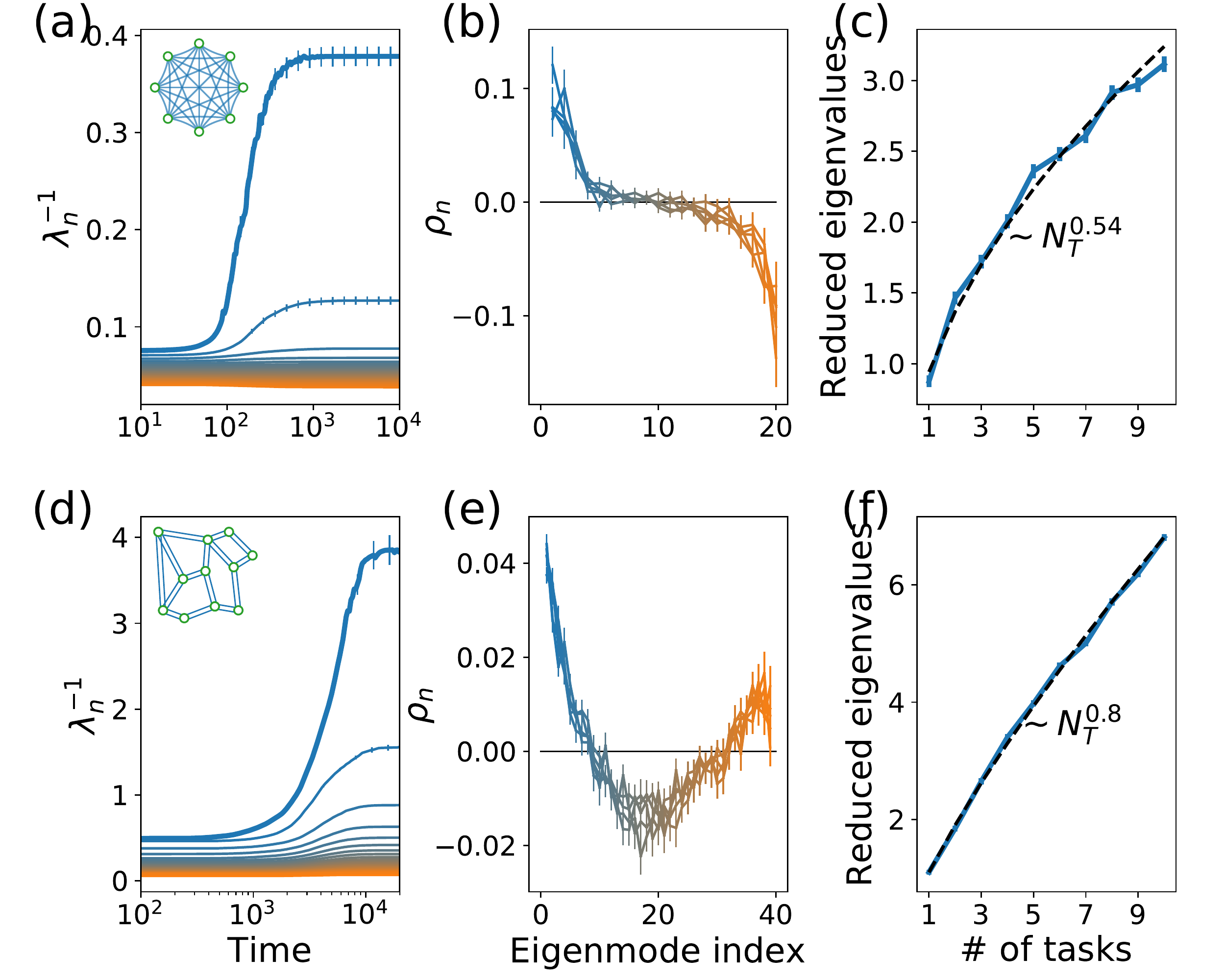}
		\caption{Hessian changes for systems learning several tasks. a) Hessian inverse eigenvalues for a fully connected linear network trained for $5$ independent tasks, with a few of the top inverse eigenvalues being modified significantly. b) Alignment of eigenvectors to the $5$ tasks shows that $\sim 3$ of the eigenmodes align with the tasks. \added{c) Varying the number of simultaneous learned tasks, we observe an increase in the number of lifted inverse eigenvalues, scaling as a power law. d-f) Similar results obtained for flow networks learning several tasks. Results are averaged over $300$ realizations of networks and tasks.}
			\label{fig:FewTasks}}
	\end{figure}
	
	%Above we saw how physical networks change when they learn a desired behavior in the small input force regime, where the physical response is described by the physical Hessian at the minimum of the native state basin. 
	Above we saw how learning, in the small input force regime, changes the Hessian and its eigenspace to accommodate a learned task, aligning an eigenmode with the task induced coordinate system, lowering the associated eigenvalue, and creating a softer mode in the physical cost landscape. In this section we extend this reasoning to physical learning of multiple tasks in the same system, whereby the Hessian changes by averaging over single task modifications, as in Eq.~\ref{eq:3a3}. We thus expect the Hessian eigenvectors to align with the different tasks. Since these tasks are in general independent from one another, training is expected to result in aligning more than one of the Hessian eigenmodes. To verify this reasoning, we train fully connected node networks with $N=20$ nodes and flow networks with $N=40$ nodes to simultaneously satisfy five randomly sampled tasks (Fig~\ref{fig:FewTasks}abde, all results averaged over $300$ realizations). In all cases, the tasks were learned well, resulting in vanishing error.
	
	We see in Fig.~\ref{fig:FewTasks}a that several inverse eigenvalues are significantly raised in these networks. In the node networks, $3$ inverse eigenvalues are raised, and we also observe that $3$ eigenmodes align (by dot product) with the tasks (Fig.~\ref{fig:FewTasks}b), having positive alignment values $\rho_n$. In flow networks we see that all inverse eigenvalues increase, with larger effects at the bottom of the spectrum, namely for the lowest eigenvalues or largest inverse eigenvalues  (Fig.~\ref{fig:FewTasks}d). Furthermore, the eigenmodes associated with the top $~5$ tend to align with the input forces (Fig.~\ref{fig:FewTasks}e). \added{It is evident from these simulations that the number of raised inverse eigenvalues (and associated aligning vectors) may be smaller than the number of tasks. To test this, we trained these types of networks simultaneously for varying numbers of tasks in the range $1-10$. For these simulations, we measured the number of raised inverse eigenvalues, defined as the number of eigenvalues which were raised by at least $10\%$ compared to their initial value. Given this definition, we find that the number of raised inverse eigenvalues scales as a power law in the number of tasks for both types of networks, but with different exponents (Fig.~\ref{fig:FewTasks}cf). While it is clear the number of modes `used' by the system for learning increases when more tasks are learned, this increase is definition dependent, and we leave more precise study and quantification of this effect to future study.}
	
\begin{figure}
		\includegraphics[width=0.95\linewidth]{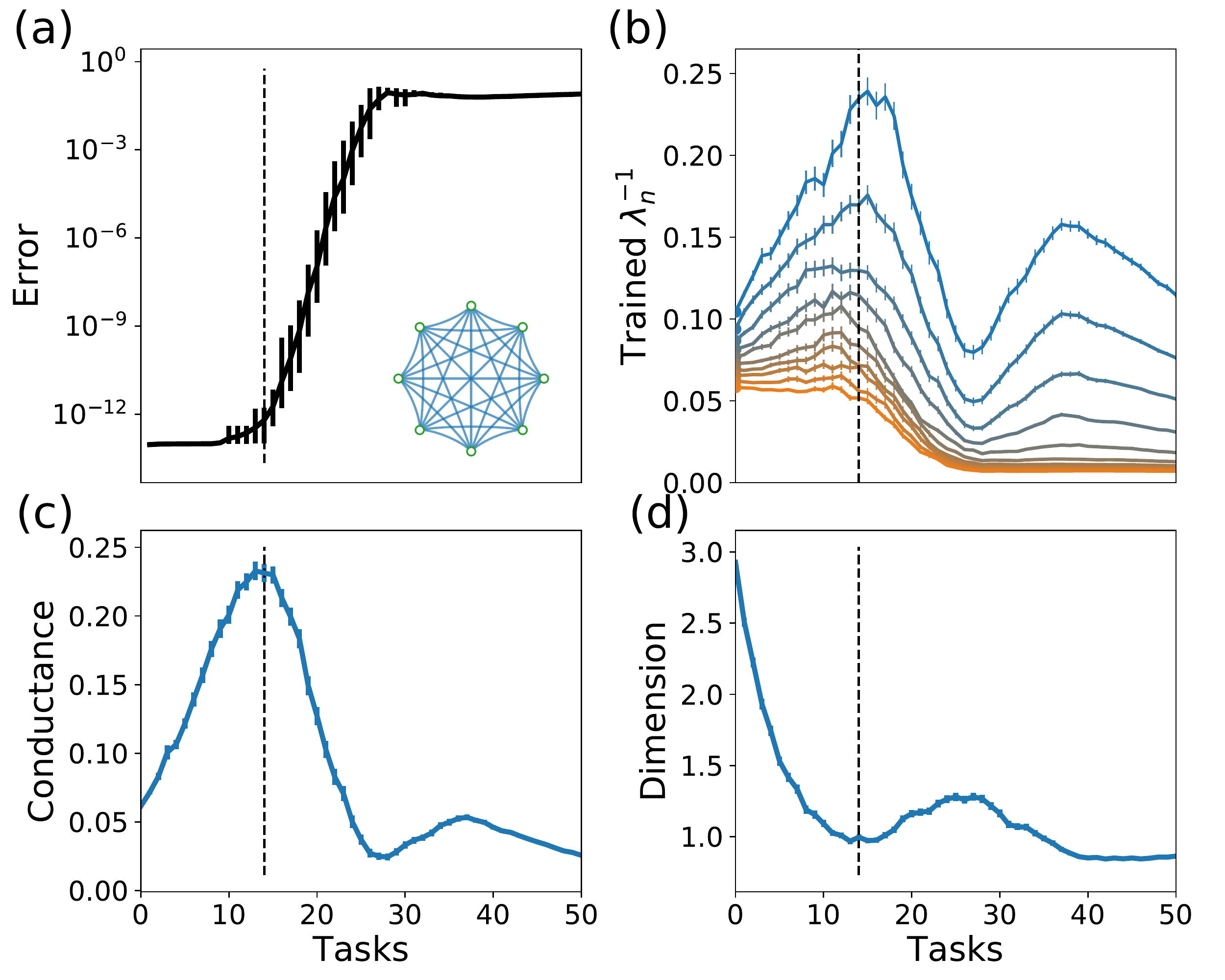}
		\caption{Training physical networks for multiple tasks. a) Training error remains low up until a capacity of tasks is reached. b) Inverse eigenvalues of a network trained far beyond capacity tend to decrease. c) Effective conductance as a function of the number of tasks - systems trained beyond capacity become less conductive (stiffer), as opposed to systems successfully trained below capacity. d) Physical response dimension remains low regardless of the number of learned tasks. Results averaged over $400$ realizations of networks and tasks.
			\label{fig:ManyTasks}}
	\end{figure}

	It is well-known that a system cannot be trained to perform too many tasks simultaneously; physical and computational learning models have a \emph{capacity} $M_C$ for trained tasks. Trying to learn beyond capacity results in failure, where the system cannot successfully perform all of the desired tasks~\cite{stern2023learning}. The capacity of simple learning models typically scales at best with the number of learning degrees of freedom (see Appendix D, where we argue our physical networks have a capacity that scales linearly with the number of learning degrees of freedom). This has been established for flow networks, where the number of output nodes that can be trained to respond to a single input is sublinear~\cite{rocks2019limits} in the total number of nodes, but can be raised to linear scaling~\cite{ruiz2019tuning} by avoiding frustration by tuning outputs in order of increasing distance from the source. We observe this finite capacity when training our models for multiple tasks (Fig~\ref{fig:ManyTasks}a). Thus, we studied the physical effects of training beyond capacity.
	%We are however more interested in studying the physical effects of the learning process. 
	
	%Why are the eigenvalues of a system trained beyond capacity increasing, in contrast with training below capacity? 
	\added{Consider again the learning cost function for a linearized task $c^{(r)} \approx \vec{A}_{r}^T(\vec{x}^F_r-\vec{x}^0) - B_r$. Training for such a task yields output nudging forces as in Eq.~\ref{eq:2.a4}. We can write down the average change in an eigenvalue as in Eq.~\ref{eq:3a6}, but retain the second term of the output force that we previously neglected, which we will see comes to dominate. The total change in the eigenvalues is then computed as the sum of the changes due to the entire set of $n_T$ tasks.
		\begin{equation}
			\begin{aligned}
				\delta \lambda_n^{(r)} &= \alpha [-B + \sum_c\frac{f_c^{(r)} a_c^{(r)}}{\lambda_c}] \sum_d Y_{dn}\frac{f_d^{(r)} a_d^{(r)}}{\lambda_d^2} \\
				\delta \lambda_n  &= n_T^{-1}\sum_r  \delta \lambda_n^{(r)} 
			\end{aligned}
			\label{eq:4.2}
		\end{equation}
		Here, $r$ is a sum on tasks, and the first line is the same as Eq.~\ref{eq:3a6} except we have now kept both terms in the square brackets.}
	
	%If we sum Eq.~\ref{eq:4.3} over many random independent tasks $\vec{F}_{r}, \vec{A}_{r}, B_{r}$, the second term in this sum averages to zero. However, as $(\vec{F}_r\vec{F}_r^T)$, $(\vec{A}_r\vec{A}_r^T)$ and $H^{-1}$ are square, positive semi-definite matrices, the first term produces a non-vanishing contribution.  Now we observe from Eq.~\ref{eq:3a6} that the change in the Hessian eigenvalues during learning is controlled by the quantity $ v^T_a \vec{F} (\vec{F}^O)^T v_a$ where we are projecting the outer product of the input and output forces onto the eigenvectors.  
	
	\added{To ask how the eigenvalues change for a typical set of tasks, we can take the components of the input forces $F_r$ and constraint elements $A_r$ to be independently sampled from zero mean, unit variance Gaussian distributions $\mathcal{N}(0,1)$ (See Appendix A). Thus, the input and output forces are random relative to any given eigenmode $v_a$ and to each other. The first term proportional to $B$ vanishes when summed over the ensemble of such forces, and the second term only has a finite contribution when we consider the same component $c=d$.
		\begin{equation}
			\begin{aligned}
				\delta \lambda_n &\approx \alpha n_T^{-1}\sum_{d,r} \frac{Y_{dn}}{\lambda_d^3} (f^{(r)}_d)^2 (a^{(r)}_d)^2\approx \alpha \sum_{d} \frac{Y_{dn}}{\lambda_d^3} \\
				\delta \lambda_n^{-1}  &\approx - \frac{\alpha}{\lambda_n^2} \sum_{d} \frac{Y_{dn}}{\lambda_d^3}
			\end{aligned}
			\label{eq:4.3}
		\end{equation}
		The second expression in the first line has taken the sum over the tasks.  We assumed that there are a large number of tasks ($n_T \gg N^2$) and that they are drawn from a random ensemble (every component of the vectors $\vec{F}$ and $\vec{A}$ for each task are drawn independently from the unit sphere).  In this limit, the learning problem is necessarily over capacity.}
	
	We see that all eigenvalues are expected to receive positive adjustments that scale with the inverse eigenvalues raised to the third power.  Thus, learning will make the system less conductive (stiffer). Below capacity, the Hessian eigenspace can align with the learned task(s), so that the lower eigenvalues can be effectively reduced. Above capacity, the Hessian cannot align with all the tasks, such that trying to learn some of them will tend to increase all eigenvalues.  Eq.~\ref{eq:4.3} gives the limiting behavior for averaging over an infinite number of tasks, but training over capacity is likely to cause an upward shift for most eigenvalues during learning.
	
	This reasoning is supported in Fig.~\ref{fig:ManyTasks}, where we train a fully connected node networks with $N=10$ nodes for an increasing number of simultaneous tasks (all results averaged over $400$ realizations of the network and tasks). In Fig.~\ref{fig:ManyTasks}a, we plot the error after training as a function of the task number, demonstrating the finite capacity of the network (dashed line, $n_T \sim 13$), defined here as the threshold above which some of the trained networks are unable to reduce the error to zero on all tasks.  In other words, we are defining capacity as the threshold in the number of tasks over which the learning process begins to fail.  Fig.~\ref{fig:ManyTasks}b shows the eigenvalues of the trained networks. As observed above, up to capacity the top inverse eigenvalues tend to increase by learning. However, approaching capacity and beyond, the inverse eigenvalues are shifted down during learning, as suggested by Eq.~\ref{eq:4.3}. %Examining the eigenmodes of systems trained beyond capacity at $n_T=30$ (Fig.~\ref{fig:ManyTasks}c), we see that the bottom eigenmodes no longer align with the trained tasks, but rather the top eigenmodes (weakly) align with them.
	
	In Fig.~\ref{fig:ManyTasks}c-d, we plot the effective conductance $\bar{g}$ and the physical response dimension for systems trained for multiple tasks. The effective conductance is maximal (dimension minimal) when the number of tasks matches the capacity for simultaneous tasks as defined above ($n_T\sim 13$).  The conductance  declines if we try to train additional tasks, and reaches a minimally conductive (stiffer) state when trained beyond capacity.  In contrast, the effective dimension of the network responses remains low even beyond capacity. These results can be explained via Eqs.~\ref{eq:3b.3},\ref{eq:3c.2} by the observation that all inverse eigenvalues of the Hessian tend to decrease when training beyond capacity, not only the top inverse eigenvalues.  The effective dimension is controlled by the {\it relative} values of the inverse eigenvalues, and hence decreasing all of them together does not change the dimension much. \added{A peculiar feature emerges in this data: corresponding to a bump in the measures of Fig.~\ref{fig:ManyTasks}c-d when trained somewhat above capacity (with $\sim 35$ simultaneous tasks). The reason for the appearance of this feature is unclear and requires further study, but it may be related to the known tendency of linear learning algorithms to have an optimal model complexity in the under-parameterized (too many tasks) regime, often referred to as the bias-variance trade-off~\cite{rocks2022memorizing}.}
	
	Finally, we note that a physical system trained for multiple tasks can naturally be subject to noise during training: If the system learns in response to every observed task independently, the order in which tasks are presented can affect the learning process. Sampling training tasks randomly gives rise to a stochastic gradient descent-like algorithm~\cite{chaudhari2018stochastic}, which can potentially affect the physical properties of the trained system as well. Moreover, physical learning is likely subject to other sources of noise, e.g., drift in the learning degrees of freedom, that affect the precision of the learning rules implemented by the system. In appendix E we study physical learning in such noisy conditions, and find that mild noise conditions, where learning is still possible, do not modify the physical effects discussed; the learning networks still find noisy solutions with high conductance and low physical dimension.

	\section{Discussion}
	
	%Learning is often studied in a regime where the goal of training is to restructure the landscape to have multiple specific minima (e.g. in Hopfield networks). In the language of our physical networks, this is the regime of large input signals, capable of driving the system between basins in its physical landscape. 
	In artificial neural networks, learning corresponds to traversing a path within the learning cost landscape that descends to a minimum. In physical learning systems, the physical cost landscape also changes as the system approaches a minimum of the learning cost function, affecting the trajectory in the learning landscape. We chose to focus  on the linear regime of small input signals where we can obtain the greatest insight into how learning restructures the physical landscape. We showed that networks can learn complex tasks ranging from digit recognition~\cite{stern2021supervised} to allostery already in this weak input regime.  The structure of the response space is characterized by the spectrum of the physical Hessian around the minimum of the physical cost function, and the structure of the Hessian eigenmodes relative to the learned tasks.
	
	While previous works demonstrated that training lowers low eigenvalues in the linear regime~\cite{tlusty2017physical, yan2018principles}, we have now traced the evolution of the eigenvalues and eigenvectors of the physical Hessian during the learning process.  We find that physical systems learning in the linear regime develop distinctive physical effects including strong responses to random inputs, as well as low-dimensional mechanical responses. This is remarkable as there are, in principle, many possible networks that satisfy a desired function without having this distinctive low-dimensional nature. If the number of tasks networks are trained for is below capacity (See Appendix~D for more details on this capacity) they can essentially learn them all perfectly and the system can find multiple solutions.  We showed that below capacity, networks physically evolve during training by becoming more conductive (less stiff in the case of elastic networks),  lowering their effective response dimension, and aligning their eigenmodes  with the learned tasks. In contrast, networks trained above capacity fail to learn and become \emph{less} conductive (stiffer), but still maintain a relatively low effective response dimension.  Thus an anomalously low network response dimension is a signature of learning, both below and above capacity. Our finding that training beyond capacity stiffens a physical system suggests a simple method of avoiding overtraining: as trainers add tasks to a network, they should test its response dynamics to random forces, stopping when the stiffness begins to increase.
	
	Our results suggest that low dimensionality is a generic outcome of learning in physical networks, possibly shedding light on open questions of why networks of neurons in the brain manifest surprisingly low dimensional response spaces. The generality of our approach further suggests a tool for analysing seemingly random physical networks to discover whether, and for what purpose, they are trained. Specifically, an experiment could measure system responses to small perturbations and then our results suggest that these responses correlate to the tasks that the system was trained for.  In other words, our results justify the naive intuition that the more responsive dimensions of a complex system encode its learned behaviors. Such tools can be useful in understanding newly discovered trained or evolved networks regardless of the specific details guiding their physical responses to perturbations.
	
	Our results were obtained in a scenario where the training data and the learning dynamics are noiseless.  We tested that our findings are robust to the addition of Gaussian noise to the learning rule with a magnitude small compared to the learning rate (Appendix E). We also observe that introducing stochasticity in the selection of the order of training examples, or the order in which edges are modified, does not change our results. It would be interesting to extend our results to investigating noisier conditions when learning becomes challenging.
	
	Finally, we note that we have shown that learning in the linear regime already has a remarkable phenomenology that can be analyzed powerfully. It would be very interesting to extend our results to nonlinear situations where the input and output forces are large. In this case, for example, the free state resulting from the application of an input may lie in a different basin of the physical cost function than the native state of the network in the absence of inputs.  Other learning mechanisms can then come into play, such as the shifting of basins so that the minima themselves align with desired behaviors. In these cases, the learning rule may be uncorrelated with the native state, so that learning would not necessarily create soft modes lowering the effective response dimension~\cite{rouviere2021emergence}.  However, the ubiquitous appearance of low dimensional response manifolds in systems that learn (see, e.g., ~\cite{recanatesi2021predictive}) suggests some of the findings might extend, perhaps in a different form, to physical learning with large inputs that explore multiple basins in the physical landscape.

	\subsection*{Acknowledgments}
	We thank Eric Shea-Brown, Giulio Biroli, Eric Rouviere and Sam Dillavou for insightful discussions. This research was supported by DOE Basic Energy Sciences through grant DE-SC0020963 (MS).  VB and MS were also supported by NIH CRCNS grant 1R01MH125544-01 and NSF grant CISE 2212519.  The Flatiron Institute is a division of the Simons Foundation, which provided support via Investigator Award \# 327939 (AJL). 
	
	\appendix
	
	%\section{Physical learning rules}

	\section{Learning tasks (allostery, regression, classification)}
	
	In this work we simulate learning in physical networks with local, physically realizable learning rules, approximating the gradient of a learning cost function. The learning rules themselves are explored in the main text. Here we discuss some cost functions the networks can optimize, i.e., what tasks the networks can learn to perform. We describe the prototypical tasks used in the main text to explore the physical effects of learning. We then show that the physical effects are similar regardless of the type of task the network is trained to perform.
	%that our results are broadly applicable to different types of tasks; the physical effects are similar regardless of what type of task the network is trained to perform.
	
	Each  task explored for fully connected node networks in the main text is chosen as follows. An input force $\vec{F}_{r}$ is randomly generated component by component from a zero mean, unit variance Gaussian distribution $\sim\mathcal{N}(0,1)$. Then an independent linear constraint $\vec{A}_{r}$ is also generated component by component from Gaussian distribution $\sim\mathcal{N}(0,1)$ along with a uniformly distributed scale parameter $B_r\sim U(0.2,1)$ in the range $0.2-1$. The finite values of $B_r$ while the forces have zero mean places us in the weak input force regime discussed in the main text. The linear constraint the network is trained to satisfy is then
	\begin{equation}
		\begin{aligned}
			0= c^{(r)} = \vec{A}_rH^{-1}\vec{F}_r - B_r
		\end{aligned}
		\label{eq:A1}.
	\end{equation}
	The associated cost function minimized by learning is $C^{(r)}=\frac{1}{2}(c^{(r)})^2$. The results shown in the main text for fully connected node networks are based on optimizing such cost functions. In the linear response regime, these tasks form a basis for any desired functionality. 
	
\begin{figure}
		\includegraphics[width=0.95\linewidth]{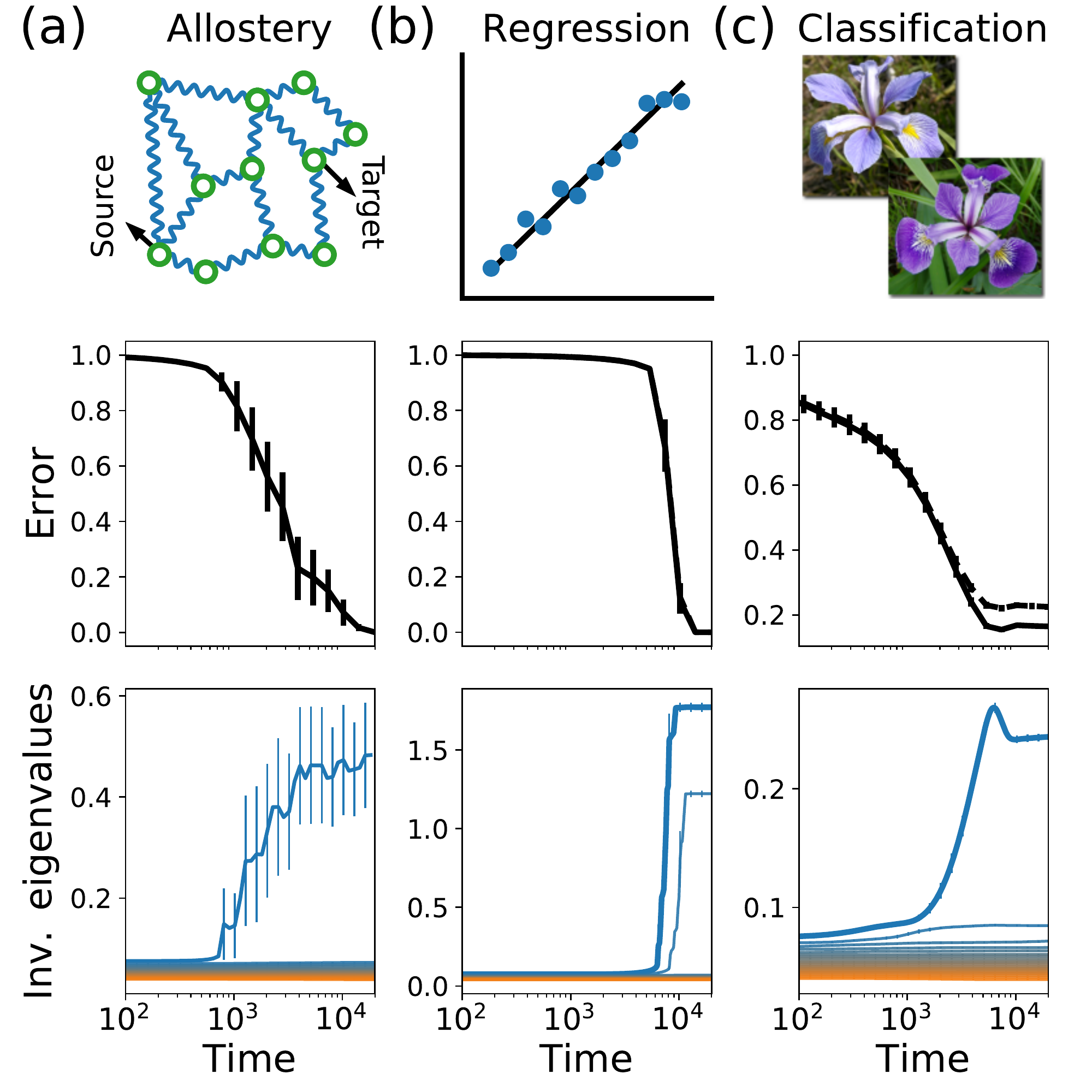}
		\caption{Training networks for different types of tasks. a) Allosteric tasks imply desired long range source-target relations. Here we show how networks can be trained for a single allosteric task involving one input and one output. Errors are shown in the middle row, and the dynamics of the inverse Hessian eigenvalues on the bottom row. b) In regression tasks, the output units of a network are trained to recover a particular linear relation with the input units. Here we train networks for a set of two equations with two variables. c) Networks are trained to classify the Iris dataset~\cite{Iris} with $4$ inputs and $3$ classes (Iris species), using $10$ out of $50$ samples per flower for training. It does so by minimizing the cross entropy error, showing full line for the training error and dashed line for the test error. For all these tasks the physical learning phenomenology is similar; learning creates physical soft modes (results averaged over $10$ networks).
			\label{fig:SI-Tasks}}
	\end{figure}

	For difference networks, like  the flow networks and mechanical networks studied here, these kind of tasks are readily learned --  networks can often learn to satisfy such constraints with small modifications of the learning degrees of freedom, and hence small changes in the physical properties. To more clearly reveal the physical effects of learning, we challenged the difference networks to learn more difficult functions.  The basic task we chose is {\it allostery}, which requires large target responses far away from source perturbations, a phenomenon previously observed in biological and mechanical networks~\cite{rocks2017designing} (see schematic in Fig.~\ref{fig:SI-Tasks}a). For each task, a random source node is chosen, and a force of magnitude $F_{r}=1$ is applied as input at that node (in the positive direction for flow nets, and in some random direction in $2D$ for mechanical spring networks). Then, a random target node $o$ is chosen, and the allosteric task is defined such that the response at that node has a finite amplitude $\vert(\vec{x}^F-\vec{x}^0)_o \vert = B$ ($B=0.3$ for flow networks, $B=0.5$, and a random response direction is chosen for the $2D$ mechanical spring networks). When multiple simultaneous tasks are considered, we select multiple pairs of sources and targets, and apply such constraints to each pair. These allosteric tasks are more challenging for difference networks to learn, so that learning produces significant modifications to the network, and the physical effects discussed in the paper can be observed readily.
	
	To show that the  physical effects of learning that we discussed are generic in the linear response regime, not only in terms of the physical network, but also in terms of the task(s), we train fully connected node networks for various types of tasks inspired by biology and computational machine learning. First, we train $N=20$ networks on allosteric tasks similar to those defined for difference networks, with results shown in Fig.~\ref{fig:SI-Tasks}a (averaged over 100 realizations of networks and allosteric tasks). We find that training consistently reduces the error by orders of magnitude, and that the eigenvalues behave similarly to what we have seen earlier in Fig.~\ref{fig:Hessian}a.
	
	Other tasks of interest in supervised machine learning include regression, where the output of a network learns to recover some relation to the inputs based on training data. We train $N=20$ fully connected linear networks to compute the results of a set of two linear equations. To do so, we randomly choose two input nodes $i_1,i_2$ and two output nodes $o_1,o_2$. Input forces correspond to the independent variables in the equations $F_{i_1}, F_{i_2}$, and the system is trained such that the response in the output nodes relates to the input forces as 
	\begin{equation}
		\begin{aligned}
			\begin{pmatrix}
				0.5 & 0.2\\
				-0.1 & 0.7
			\end{pmatrix}
			F_i=(x^F-x^0)_o
		\end{aligned}
		\label{eq:A2}.
	\end{equation}
	This is done by applying the learning rule of Eq.~\ref{eq:2.a7} for $50$ random training forces, sampled from a Gaussian distribution as before. We further sample $100$ \emph{test} forces to measure the regression performance on data that is not used for training. Results are shown in Fig.~\ref{fig:SI-Tasks}b (averaged over 10 networks and training/test sets). The networks generally succeed in learning the desired input-output relation, decreasing the MSE cost by many orders of magnitude for both the training set (solid line) and test set (dashed line). More importantly, we track the eigenspace of the Hessian during learning, and find that two eigenvalues are strongly reduced (inverse eigenvalues increased) when the network learns these two relations. Furthermore, the bottom two eigenmodes strongly align with the training set responses. This suggests the physical effects of learning are similar to the case of training for several tasks, discussed in the main text.
	
	Finally, we train networks for classification, where the system learns to assign labels to inputs.  Networks with $N=20$ nodes are trained to classify the iris dataset~\cite{Iris}, where the inputs correspond to $4$ measurements of $3$ species of iris specimens. For each species of iris we choose $10$ specimens as a training set and the $40$ others as the test set.  The Iris measurements are applied as forces at four randomly chosen input nodes. A further three output nodes are chosen to correspond to the iris species. Since this is a discrete task (iris specimens have discrete labels), the network `selects' a label by having the largest response at the associated output node. Networks are trained by minimizing a cross entropy cost function that is more appropriate for discrete classification~\cite{scellier2021deep}. In essence, the modified learning rule is the same as Eq.~\ref{eq:2.a7} for specimens that are not classified correctly, while the output force vanishes for specimens that are classified correctly. 
	
	The results are shown in Fig.~\ref{fig:SI-Tasks}c (averaged over $10$ realizations of networks and choices of training sets). As before, learning generally succeeds, significantly reducing the cross entropy error for both the training set (full line) and test set (dashed line). In terms of classification accuracy, the trained networks reach $100\%$ training accuracy and $95\%$ test accuracy. More importantly, the effects of learning on the physical network itself are again similar to the previous cases. The lowest Hessian eigenvalues are significantly lowered by learning (inverse eigenvalues increased), reducing the physical dimension and increasing the effective conductance of the system. These results suggest that in the linear response regime the effects of learning on physical systems are generic and do not depend on the desired function.

	\section{\added{Eigensystem dynamics for learning in general linear physical networks}}
	
	\added{In this section we derive in more detail the main results for learning in linear physical networks. We show that when these networks learn to perform tasks characterized by weak input forces (compared to the desired output and initial curvature of the energy function), certain eigenvalues of the physical Hessian are decreased and associated eigenmodes align with the learned task. }
	
	\added{Assume a linear physical system whose energy is given by Eq.~\ref{eq:2.1}. For convenience, let the native state define the origin $\vec{x}^0=0$. To discuss a generic linear system, we consider a family of possible physical Hessians at the native state, given by}
	
\begin{equation}
			\begin{aligned}
				H_{ab} &= \frac{1}{2}\sum_i \phi_i(w_i) [L_{ai} R_{ib} + R^T_{ai}L^T_{ib}]\equiv \sum_i H_{ab}^i.
			\end{aligned}
			\label{eq:GenHess}
	\end{equation}
	
	\added{This Hessian can be understood as follows. There exists a set of learning degrees of freedom $w_i$, each of which is potentially subject to a non-linear transformation $\phi_i(w_i)$. These learning degrees of freedom are transported into the components of the physical Hessian $H_{ab}$ by means of a left matrix $L_{ai}$ and a right matrix $R_{ib}$, which determine how the different physical degrees of freedom $x_a$ interact through the learning parameters. For this to be a sensible physical Hessian, note that we must require that it is symmetric and positive definite. The symmetry requirement is always satisfied by the sum of the two terms in Eq.~\ref{eq:GenHess}, but one should be more careful about the Hessian being positive definite (with only positive eigenvalues). Also note that the matrices $L,R$ describe the `geometry' of the network and are fixed during learning; only the learning degrees of freedom $w_i$ are modified. The specific physical networks discussed in the main text are given by special choices for the $L,R$ matrices and a linear relation $\phi_i(w_i)=w_i$. }
	
	\added{For the flow and mechanical systems described in the main text, the left and right matrices are the same with $R=L^T=\Delta$, the incidence matrix (Eq.~\ref{eq:Widentification}). That is not the case for the node network (Eq.~\ref{eq:2.a9}). We chose a smart representation of the learning degrees of freedom in a square matrix, with the left and right matrices the same, $H_{ab}(w_{ab}) = \frac{1}{2}\sum_{ij}[\delta_{ai}w_{ij}\delta_{jb} + \delta_{bi}w_{ij}\delta_{ja}]$, a choice which greatly simplifies calculations. However, if we had represented the learning degrees of freedom as a vector $w_i$, the Hessian for the node network would have different left and right matrices, corresponding to the way in which every element of the vector $w_i$ should be placed in the correct component of $H_{ab}$.}
	
	%[ Put a short paragraph reminding of the node nework in the main text as an example, and then also write it in the representation where the edge parameters are in a vector and say that in this case L and R will be different from each other.]
	
	\added{We previously discussed how such a network can learn a task in the context of weak input forces. For a given linear constraint the network must satisfy, a learning step is performed by clamping with a particular output force (Eq.~\ref{eq:4.2})
		\begin{equation}
			\begin{aligned}
				\vec{F}^O= -(\vec{A}\vec{A}^T) H^{-1} \vec{F} + B\vec{A} \approx B\vec{A},
			\end{aligned}
			\label{eq:GenFO}
		\end{equation}
		where we assumed the input force $\vec{F}$ is weak, or conversely that the physical Hessian $H$ is initially stiff (with low inverse eigenvalues) compared to the desired output scale given by $B>0$. As discussed in the main text, this is a typical case for training realistic physical networks where the input and output sectors are generally not correlated. We make the weak input assumption so that the output force remains essentially fixed throughout training (see explanation in main text). We now use the inherent coordinate system of the Hessian, specified by a matrix $v$ with rows that are eigenvectors so that $H=v^T \Lambda v$ to rotate the input and output force vectors into these natural coordinates, $f_a\equiv v_{ab}F_b$ and $f_a^O\equiv v_{ab}F_b^O\approx B v_{ab}A_b^O\equiv Ba_a$.}
	
	\added{Our derivation of the physical learning dynamics in Eq.~\ref{eq:2.a8} is valid for any Hessian $H$, so we can write (setting the learning rate $\alpha=1$) 
		\begin{equation}
			\begin{aligned}
				\delta w_i &\approx - \vec{F}^T \, H^{-1} \frac{\partial H}{\partial w_i}  H^{-1} \vec{F}^O  \\
				&\approx - B \vec{f}^T \Lambda^{-1} v \frac{\partial H}{\partial w_i} v^T \Lambda^{-1} \vec{a}  \\
				&= - \frac{B}{2} \vec{f}^T \Lambda^{-1}  v[\phi_i' (LR + R^T L^T)_i ]v^T \Lambda^{-1} \vec{a}  \\
				%&\equiv -\frac{B}{2}(\vec{\frac{f}{\lambda}})^T v[ \phi_i' [LR + R^T L^T]^i]v^T  (\vec{\frac{a}{\lambda}})  \\
				&\equiv -B\sum_{cd} \frac{f_c}{\lambda_c} M^i_{cd}  \frac{a_d}{\lambda_d}
			\end{aligned}
			\label{eq:GenDW}
		\end{equation}
		where we defined a set of square symmetric matrices, $M_{ab}^i \equiv =\sum_{cd} v_{ac} \frac{\partial H_{cd}}{\partial w_i} v_{db}^T$, one for each learning degree of freedom. }
	
	%where we defined vectors rotated to the natural coordinates $\vec{\frac{f}{\lambda}}, \vec{\frac{a}{\lambda}}$ and scaled by the eigenvalues of the Hessian, and a set of matrices, one for each learning degree of freedom, $M_{ab}^i \equiv \frac{1}{2} \phi_i' \sum_{cd}v_{ac} [L_{ci} R_{id} + R^T_{ci}L^T_{id}]v^T_{db}$, all square symmetric matrices. %Note that for a system with random values of the learning degrees of freedom $w_i$ and network geometries, the matrices $\partial H_{cd}^i \equiv \phi_i'[L_{ci} R_{id} + R^T_{ci}L^T_{id}]$ appear random relative to the physical Hessian $H$ and its eigenmodes $v$. Recalling that the eigenmodes of $H$ are orthonormal, that means on average, for two different modes $a\ne b$, $\langle M_{ab}^i \rangle=0$ and $\textrm{var}(M_{ab}^i) \sim \sum_a (\xi_{a}^i)^2$, where $\xi_{a}^i$ are the eigenvalues of the matrix $\partial H_{ab}^i$. However, when we look at the components of $M_{aa}^i$ associated with a particular eigenmode, their alignment dictates that $\langle M_{aa}^i \rangle\sim \sum_a \xi_a^i$ and $\textrm{var}(M_{aa}^i)\sim \sum_a (\xi_{a}^i)^2$. 

	\added{The change in each Hessian component $H_{ab}$ is given by Eq.~\ref{eq:3a1}, where we assume, as before, that the change in the native state $\vec{x}^0$ due to learning is small. Therefore we can write
		\begin{equation}
			\begin{aligned}
				\delta H_{ab} &= \sum_i \frac{\partial H_{ab}}{\partial w_i} \delta w_i  \\
				&=-B\sum_{i,cd}  \frac{\partial H_{ab}}{\partial w_i} \frac{f_c}{\lambda_c} M_{cd}^i  \frac{a_d}{\lambda_d}
			\end{aligned}
			\label{eq:GenDH}
		\end{equation}
		As in the main text, we utilize first order perturbation theory to compute the change in eigenvalues due to learning
		\begin{equation}
			\begin{aligned}
				\delta \lambda_n &\approx \sum_{ab} v_{na} \delta H_{ab} v^T_{bn}  \\
				&=-B\sum_{cd} \frac{f_c}{\lambda_c} [ \sum_{i} M_{nn}^i  M_{cd}^i]  \frac{a_d}{\lambda_d} 
			\end{aligned}
			\label{eq:GenDL}
		\end{equation}}
		
		\added{Similarly, we use perturbation theory to compute the modification of each eigenvector:
		\begin{equation}
			\begin{aligned}
				\delta v_{na} &\approx \sum_{m\ne n}\frac{\sum_{cd} v_{mc} \delta H_{cd} v^T_{dn}}{\lambda_n - \lambda_m} v_{ma}  \\
				&\approx B \sum_{m\ne n}\frac{\sum_{cd} \frac{f_c}{\lambda_c} [\sum_i M_{mn}^i  M_{cd}^i ]\frac{a_d}{\lambda_d}}{\lambda_m - \lambda_n} v_{ma}
			\end{aligned}
			\label{eq:GenDV}
		\end{equation}
		We see that the eigenvalue and eigenvector dynamics are dictated by the product of the input and output forces $\frac{f_c}{\lambda_c}, \frac{a_d}{\lambda_d}$ (in the natural Hessian frame of reference), with inner product defined by the matrix $\mathcal{M}^{cd}_{mn}\equiv \sum_i M_{mn}^i M_{cd}^i$. We will next given an argument for the structure of these matrices $\mathcal{M}_{mn}$. }
	
	%Note that for a large system with many interactions, each with a random value of the learning degrees of freedom $w_i$, the matrices $\frac{\partial H_{ab}}{\partial w_i}$ (for any $i$) appear random relative to the physical Hessian $H$ in the sense that an eigenmode of $\frac{\partial H}{\partial w_i}$, $\vec{u}_{a}^i$, and an eigenmode of $H$, $\vec{v}_{b}$, are effectively two random vectors on the surface of a high-dimensional $N$-sphere.
	
	\added{Using our model for the physical Hessian and the definitions above, we can explicitly write this matrix as
		\begin{equation}
			\begin{aligned}
				\mathcal{M}^{cd}_{mn} = \sum_i &\frac{(\phi'_i)^2}{4} [(\vec{v}_m L)_i(R \vec{v}_n)_i(\vec{v}_c L)_i(R \vec{v}_d)_i + \\
				&+ (\vec{v}_m L)_i(R \vec{v}_n)_i(R\vec{v}_c)_i (\vec{v}_d L)_i+ \\
				&+ (R \vec{v}_m)_i(\vec{v}_n L)_i(\vec{v}_v L)_i (R \vec{v}_d)_i+ \\
				&+ (R \vec{v}_m)_i(\vec{v}_n L)_i(R \vec{v}_c)_i (\vec{v}_d L)_i]
			\end{aligned}
			\label{eq:BigM}
		\end{equation}
		Here $\vec{v}_m$  denotes the eigenvector associated with the $m$th eigenvalue. Eq.~\ref{eq:BigM} is a sum over edges $i$, where each summand is a sum of four terms, themselves products of four vectors that involve the components of eigenvectors of the Hessian $\vec{v}_n$.
		%Let us now make an assumption about the interaction matrices $L,R$.
		We will assume that initial system is a random network, i.e., the interactions in the system connect randomly chosen physical degrees of freedom.  We will also assume that the rows of $L$ and columns of $R$ sum to zero on average. The latter assumption  is true exactly for every interaction of a flow or elastic network, because, as described above, the entries of $L$ and $R$ are signed depending on whether the edge flows in our out of the node degree of freedom.  Here we relax this assumption to allow for networks in which this is only true on average for each interaction. We also assume that each interaction,  or network edge contributes positively to the total energy of the system even when observed in isolation.  This is also true for flow and elastic networks.}
		
		\added{With these assumptions, we can understand the behavior of products like $(\vec{v}_m L)_i(R \vec{v}_n)_i(\vec{v}_c L)_i(\vec{v}_d R)_i$. First, sums over interactions $\sum_i (\vec{v}_m L)_i, \sum_i (\vec{v}_m R)_i$ will vanish for large systems because the sum on $i$ will add equally many plus and minus ones on average for every component of $\vec{v}_m$.
		So, in the random initial network, we can treat $\sum_i (\vec{v}_m L)_i, \sum_i (\vec{v}_m R)_i$ as zero mean random variables.
		%ConsiderTerms of the form $(\vec{v}_m L)_i, (R \vec{v}_m)_i$ are linear combinations of numbers, and sum
		%with the weights summing on average to zero. As random interactions with the elements of $L,R$ having their sign flipped are equally likely, we see that the sums over interactions $\sum_i (\vec{v}_m L)_i, \sum_i (\vec{v}_m R)_i$ tend to zero.
		Next, recall that the eigenvectors $\vec{v}_m$ are high-dimensional orthonormal vectors, resulting from large random physical networks. Therefore, their components for different vectors should be weakly correlated. In that case, for our random vectors,  $(\vec{v}_m L)_i (R \vec{v}_n)_i$ is effectively the product of two independent zero mean random variables unless $m=n$, in which case we are looking at correlated variables. Thus the sum $\sum_i (\vec{v}_m L)_i (R \vec{v}_n)_i$ can only be non-zero if we consider the same eigenvector $m=n$.  We can continue similarly for the products four quantities appear in each term of the summand of (\ref{eq:BigM}).}
		
		\added{We find that to have a finite contribution in Eq.~\ref{eq:BigM}, the indices $c,d,m,n$ must appear equal in pairs, or all be equal. All other options, where at least one index has a unique value, cause the associated component of $\mathcal{M}$ to vanish after summing on $i$. We confirmed this observation numerically for  networks with randomly chosen interactions satisfying the conditions described above. The vanishing of all these components becomes more precise as the system size, and the total number of random interactions increases. We therefore find that for large systems, only components of $\mathcal{M}$ where indices appear identical in pairs will contribute to the eigensystem dynamics. We next argue that  these remaining components are non-negative.}
		
		\added{If all the indices are equal, we have $ \mathcal{M}_{nn}^{nn} = \sum_i (\phi '_i)^2 (\vec{v}_n L)_i^2 (R\vec{v}_n)_i^2 \equiv X_{nn} > 0$. The next case of interest is when the bottom indices and top indices of $\mathcal{M}$ are equal in pairs,  $\mathcal{M}^{cc}_{nn}$. The result is given by $\mathcal{M}_{mm}^{nn} \equiv Y_{mn} = Y_{nm}$. We can show that $Y_{mn}$ is non-negative as follows. First, note that the physical system must be stable in its native state and that we have assumed that each interaction term is separately non-negative, i.e., $2E_i(\vec{x})=\phi_i\sum_{ab} x_{a}L_{ai}R_{ib}x_{b} \equiv \phi_i (\vec{x}L)_i (R\vec{x})_i \geq 0$. Now we observe that the components of $Y$ are given by:
		\begin{equation}
			\begin{aligned}
				Y_{mn} &= \sum_{i} (\phi '_i)^2 (\vec{v}_m L)_i(R \vec{v}_m)_i(\vec{v}_n L)_i(R \vec{v}_n)_i  \\
				&= 4 \sum_{i} (\frac{\phi '_i}{\phi_i})^2 E_i(\vec{v}_m)E_i(\vec{v}_n) \geq 0
			\end{aligned}
			\label{eq:BigY}
		\end{equation}}
		
		\added{The last non-vanishing case occurs when the top and bottom pairs in $\mathcal{M}^{cd}_{mn}$ are equal, but the two members of the pair differ, i.e. $m=c,n=d$ or $m=d,n=c$. In these cases we have $\mathcal{M}_{nm}^{nm} = \sum_i \frac{(\phi'_i)^2}{4} [(\vec{v}_n L)_i^2 (R\vec{v}_m )_i^2 + (\vec{v}_m L)_i^2 (R\vec{v}_n)_i^2] + \frac{1}{2} Y_{nm}  \equiv X_{nm} = X_{mn} > 0$. As above these matrix elements must be positive. }
		
		\added{We are now in position to estimate the change in the eigenvalues and eigenvectors due to one learning step. The eigenvalue change is:}
		
\begin{equation}
			\begin{aligned}
				\delta \lambda_n &\approx -B\sum_{cd} \frac{f_c}{\lambda_c} Y_{nc}\delta_{cd}  \frac{a_d}{\lambda_d}  \\
				&=-B Y_{nn} \frac{f_n a_n}{\lambda_n^2} - B\sum_{c\ne n} Y_{nc} \frac{f_c a_c}{\lambda_c^2}
			\end{aligned}
			\label{eq:GenDL2}
		\end{equation}
		
		\added{The first term reduces the eigenvalue $\lambda_n$ if the associated components of the input and output forces align ($f_n a_n>0$) or increases it otherwise. The second term tends to decrease all eigenvalues if the forces align, particularly at the low end ($\sum_n \lambda_n^{-2} f_n a_n$) or increase all of them otherwise more or less equally, since the components of $Y_{nc}$ are positive, all of the same scale, and the components of the forces are random. The result of these dynamics is that eigenvalues corresponding to aligned eigenmodes ($f_n a_n>0$) tend to decrease compared to the bulk of the eigenvalues. As for the eigenmodes, we have
		\begin{equation}
			\begin{aligned}
				\delta v_{na}  &\approx B \sum_{m\ne n}\frac{\sum_{cd} \frac{f_c}{\lambda_c} X_{mn}(\delta_{mc}\delta_{nd} + \delta_{md}\delta_{nc}) \frac{a_d}{\lambda_d}}{\lambda_m - \lambda_n} v_{ma}  \\
				&= B \sum_{m\ne n}\frac{X_{mn} (f_m a_n + f_n a_m)}{\lambda_m\lambda_n(\lambda_m - \lambda_n)} v_{ma}
			\end{aligned}
			\label{eq:GenDV2}
		\end{equation}
		which leads to a rotation of the eigenmodes. We see that this rotation  depends on their alignments with the forces $f_m a_n, f_n a_m$, as well as whether the associated eigenvalue $\lambda_n$ is low or high (because of the difference in the denominator). For the bottom eigenmode, these dynamics causes stronger alignment with the input and output forces. To see this explicitly, we can estimate the change in alignment due to learning:}
		
\begin{equation}
			\begin{aligned}
				\delta &(f_n a_n)  = \sum_a \delta v_{na} F_a a_n +  f_n \sum_a \delta v_{na} A_a \approx \\
				&\approx B \sum_{m\ne n,a}\frac{X_{mn} (f_m a_n + f_n a_m)}{\lambda_m\lambda_n(\lambda_m - \lambda_n)} v_{ma}F_a a_n +\\
				&+ B \sum_{m\ne n,a}\frac{X_{mn} (f_m a_n + f_n a_m)}{\lambda_m\lambda_n(\lambda_m - \lambda_n)} f_n v_{ma} A_a = \\
				&= B \sum_{m\ne n}\frac{X_{mn} (f_m a_n + f_n a_m)^2}{\lambda_m\lambda_n(\lambda_m - \lambda_n)} 
			\end{aligned}
			\label{eq:GenAlign2}
	\end{equation}
	
	\added{We therefore see that the alignment of eigenvector $n$ with the task, defined in terms of the product of $f_n, a_n$, the input and output forces in the natural coordinates of the Hessian, must grow for low eigenmodes and decrease for high eigenmodes. The $\lambda$ scaling in the denominator means that this process occurs preferentially in the lower eigenmodes, which tend to align with the task. Connecting this with Eq.~\ref{eq:GenDL2} above, this increased alignment of the low eigenmodes will cause them to be reduces compared to the rest more effectively (due to larger $f_n a_n$). The physical networks discussed in the main text are both special cases of this theory, explaining the phenomenology of both of them. }
	
\begin{figure}
		\includegraphics[width=0.95\linewidth]{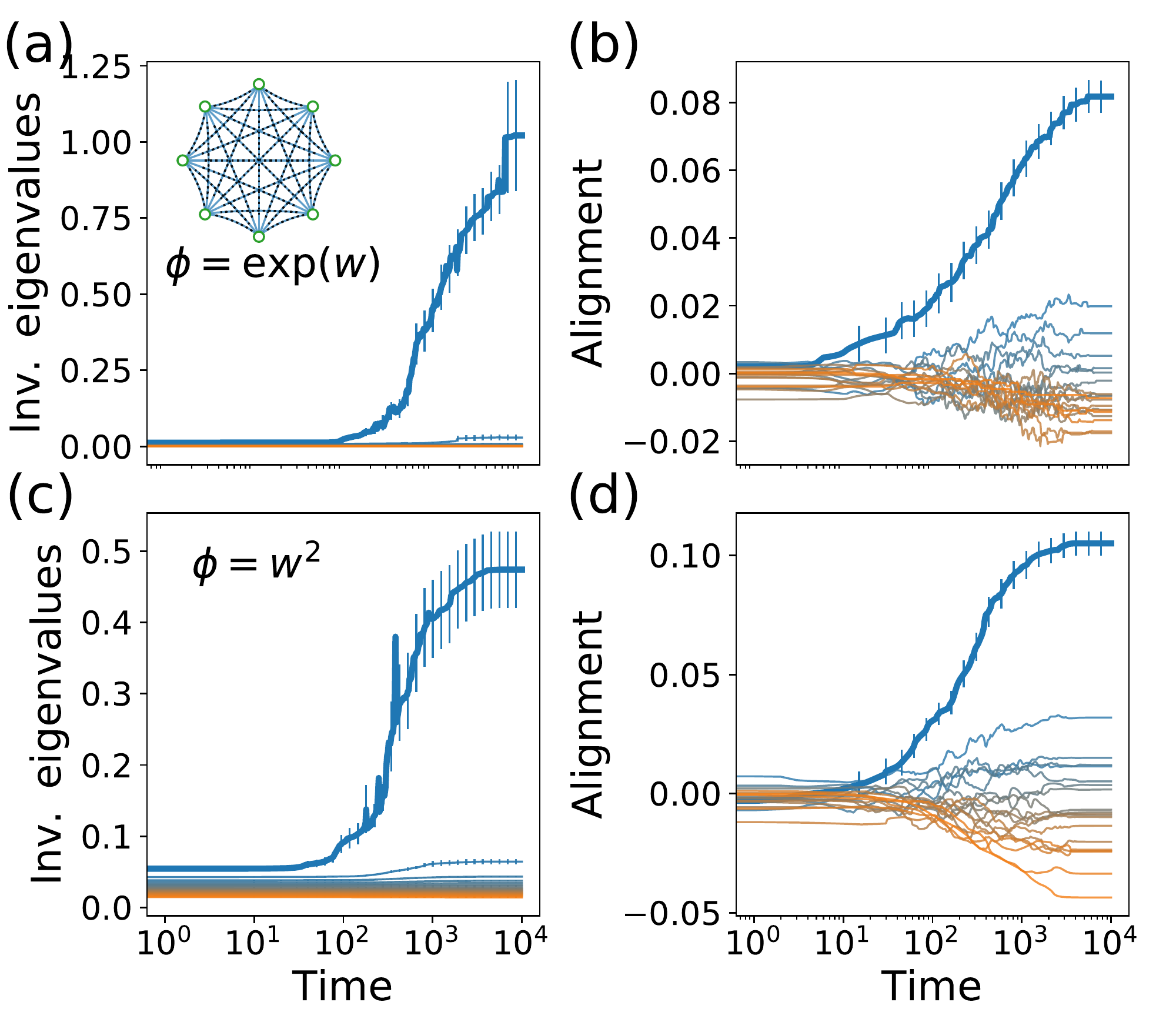}
		\caption{\added{The physical effects of learning persist when the physical Hessian is associated with edge-wise non-linearities. (a-b) Training fully connected linear networks with exponential edge-wise non-linearity $\phi(w) = \exp(w)$. Learning still causes the lifting of an inverse eigenvalue and the alignment of its associated eigenmode. (c-d) Similar results are observed for fully connected linear networks with quadratic non-linearities $\phi(w) = w^2$. Results are averaged over $150$ sets of networks and tasks.}
			\label{fig:SI-NonLinW}}
	\end{figure}

	\added{This analysis also shows that the introduction of edge-wise non-linearities does not change the overall story, since they only enter the eigensystem dynamics in squared form. To test this observation, we simulated fully connected (N=20) networks like those of Eq.~\ref{eq:2.a10}, but with nonlinear interactions of two types, $\phi_{ab}=\exp(w_{ab})$ and $\phi_{ab}=w_{ab}^2$. The results for the dynamics of the eigenvalues and mode alignment are given in Fig.~\ref{fig:SI-NonLinW}. We observe that these dynamics are qualitatively similar to those observed in the linear Hessian case of Fig.~\ref{fig:Hessian}ab: The lowest eigenvalue is reduced by learning, and the associated eigenmode aligns with the task. It is notable that in these node connected networks, the off-diagonal elements of matrix $Y_{nc}$ vanish, so that the higher eigenvalues in the spectrum tend to rise because their eigenmodes become misaligned with the task.}

	\section{Physical response dimension}
	
	In this appendix we discuss measures for the physical response dimension, in particular the effective response dimension defined in the main text. Define a large set of $M\to\infty$ random forces $\vec{F}^R_{m}$, each of which sampled from the set of normalized vectors on the $N$-sphere, where $N$ is the physical dimension of the system (number of physical degrees of freedom). The physical response of the system to these forces is given by
	
	\begin{equation}
		\begin{aligned}
			\vec{\delta}^R_{m} \equiv \vec{x}^R_{m} - \vec{x}^0 = H^{-1}\vec{F}^R_{m}
		\end{aligned}
		\label{eq:B1}.
	\end{equation}
	
	Applying the random forces, we obtain a set of $M$ physical responses $\vec{\delta}^R_{m}$. The Hessian of the native state can be decomposed to a set of non-negative eigenvalues $\lambda_a$ and associated eigenmodes $\vec{v}_a$ (whose number is equal to the number of physical degrees of freedom $N$). These eigenmodes correspond to different orthonormal ways in which the system can respond to perturbations. Thus, to estimate the response dimension, we can ``count” the number of eigenmodes participating in the response.
	
	Projecting the physical response over the Hessian eigenmodes gives the amplitude of each mode's activation due to the external force $p_{am} \equiv \vec{v}^T_{a} \vec{\delta}^R_{m}$. Summing this quantity over the eigenmodes and different random forces is self-averaging to zero because eigenmodes can be activated both positively and negatively. The physical response dimension is defined using these projections as an effective way to count the number of participating eigenmodes.  For a given random force $\vec{F}^R_m$ we have
	\begin{equation}
		\begin{aligned}
			D_m \equiv \frac{\left[\sum_a p_{am}^2\right]^2}{\sum_a p_{am}^4}
		\end{aligned}
		\label{eq:B2}.
	\end{equation}
	As discussed in the main text, this measure for the physical dimension has reasonable limits; it is $D_m=1$ if only one mode participates, and $D_m=N$ if all modes participate equally. 
	
	In view of this,
	we propose to  characterize the effective response dimension by the quantity
	\begin{equation}
		D_{{\rm eff}} 
		=
		\frac{\langle \sum_a p_{am}^2 \rangle^2}{\langle \sum_a p_{am}^4 \rangle} \, ,
		\label{eq:Deffdef}
	\end{equation}
	where the angle brackets indicate an expectation value over the ensemble of random forces.
	%, and $C$ is a constant that we will choose later.
	%As compared to the participation ration dimension (\ref{eq:PartRatDef}), we have moved the expectation value separately into the sums in the numerator and denominator. 
	Thus $D_{{\rm eff}}$ measures the effective dimension as the ratio of the square of the second moment of the projections and the fourth moment of the projections.  %This quantity also equals $1$ in the responses localize on a single eigenmode, and equals $N$ if the all the modes participate equally.

	%We can define a simpler quantity characterizing the effective response dimension by first
	We can compute $D_{{\rm eff}}$ by recalling that the inverse Hessian can be decomposed as $H^{-1}=v\Lambda^{-1} v^T$, where $\Lambda^{-1}$ is a diagonal matrix of the inverse eigenvalues $\lambda_a^{-1}$, and $v$ is a matrix whose columns are the eigenvectors:
	\begin{equation}
		\begin{aligned}
			p_{am} = \vec{v}_a\cdot \vec{\delta}^R_{m}=\vec{v}_{a} v^T\Lambda^{-1} v \vec{F}^R_{m} = \frac{(\vec{f}^R_{m})_a}{\lambda_a}
		\end{aligned}
		\label{eq:B3}.
	\end{equation}
	Here we used the fact that $\vec{v}_{a} v^T$ is a vector with zeros at all components, except a single $1$ at component $a$. Therefore, only the $a^{{\rm th}}$ component of the vector $\vec{\Lambda}_{a}^{-1}$ is nonzero, and equals $\lambda_a^{-1}$. %So, finally, we find that
	%\begin{equation}
	%P_{am} = \lambda_a^{-1} \vec{v}_a \cdot \vec{F}^R_m
	%\label{eq:pamdef}
	%\end{equation}
	
	As we saw in the main text, the set of eigenvectors $\vec{v}_a$ form a fixed orthonormal basis in $N$ dimensions, while the forces $\vec{F}^R_{m}$ are random $N$-dimensional unit vectors.  Thus high dimensional systems ($N\gg 1$) the components $\vec{f}^R_{m}$ will be  distributed according to a zero mean Gaussian with variance $1/N$, $\mathcal{N}(0, N^{-1})$.   Averaged over the ensemble of random forces, the second and fourth moments of $p_{am}$ are given by
	\begin{equation}
		\langle p_{am}^2 \rangle = \frac{1}{N} \lambda_a^{-2}
		~~~~;~~~~
		\langle p_{am}^4 \rangle = \frac{3}{N^2} \lambda_a^{-4}
		\label{eq:Pmoments}
	\end{equation}
	where we used the Eq.~\ref{eq:B3} and the standard moments of a Gaussian distribution with variance $1/N$.

	Thus,  the effective dimension in (\ref{eq:Deffdef}) is
	\begin{equation}
		D_{{\rm eff}} = \frac{(\sum_a \lambda_a^{-2})^2}{3 \sum_a \lambda_a^{-4}} \, .
	\end{equation}
	If one eigenvalue is very small and thus dominant, the effective dimension in response to random forces is $1/3$ reflecting that fact that many forces will not drive the system much at all.

	\begin{figure}
		\includegraphics[width=0.95\linewidth]{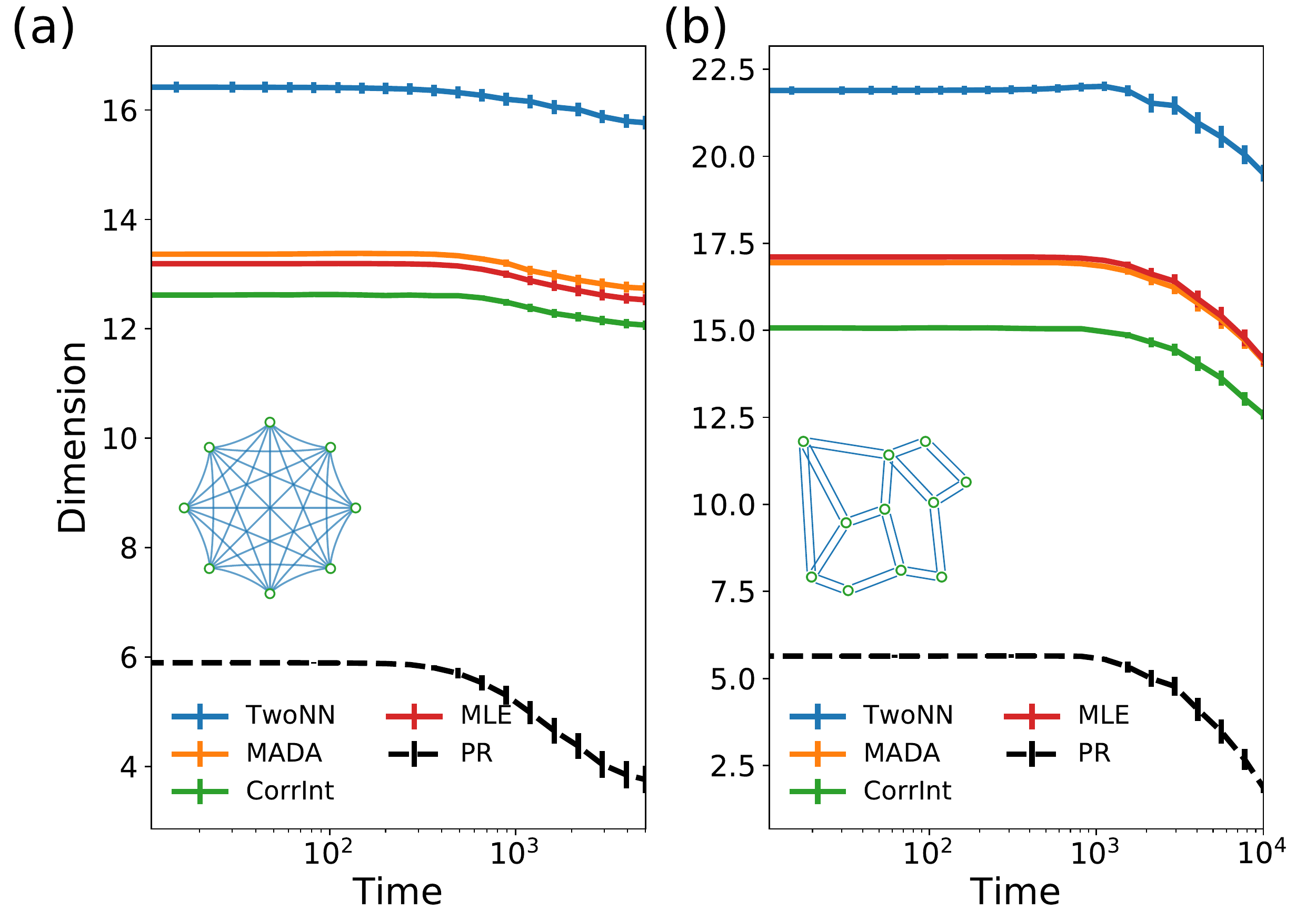}
		\caption{Different measures of the physical dimension. a) While in the main text we mainly used a measure derived from the participation ratio dimension to discuss dimensional reduction in the physical response, multiple other measures for the intrinsic dimension of the response manifold show the same result. Here we train fully connected linear networks and show how the physical dimension changes during training for several machine-learning-inspired measures. b) Similar results are found when training flow networks - physical dimension is reduced for all methods used to measure it.
			\label{fig:SI-Dimension}}
	\end{figure}
	
	While in the main text we discussed $D_{\rm eff}$ as a measure of the physical response dimension of the system, there are several other measures of the intrinsic dimension of manifolds inspired by machine learning.  We tested that our key qualitative results are independent of the choice of the measure of dimension. To do this we randomly select $500$ normalized forces and applied them to systems during training (either fully connected linear networks or flow networks). We used the resulting responses  to estimate the physical dimension using different methods: Manifold-adaptive dimension estimation (MADA)~\cite{farahmand2007manifold}, Correlation dimension (CORR)~\cite{grassberger2004measuring}, Maximum likelihood estimate (MLE)~\cite{hill1975simple} and the TwoNN algorithm~\cite{NEURIPS2019_cfcce062}. Fig.~\ref{fig:SI-Dimension} shows that the physical dimension is decreased during learning in physical systems regardless of the chosen method for dimension estimation.

	\section{Learning capacity}
	
	In the main text we discussed how physical systems are able to learn multiple tasks up to a finite capacity. Here we argue that for systems trained in the linear response regime, this learning capacity is linear in the number of learning degrees of freedom $N_w$, itself at most quadratic in the system size $N$.
	
	For small input forces in the linear response regime, any learning cost function can be expressed as a sum of linear constraints, relating response of the physical degrees of freedom $(\vec{x}^F_r-\vec{x}^0)$ to an input force $\vec{F}_{r}$:
	\begin{equation}
		\begin{aligned}
			c^{(r)}&=\vec{A}_{r}(\vec{x}^F_r-\vec{x}^0) - B_r=\\
			&=\vec{A}_{r}H^{-1}\vec{F}_{r} - B_r
		\end{aligned}
		\label{eq:D1}
	\end{equation}
	Successful learning  means that $c^{(r)}=0$, or that for every task $r$ the system solves a linear equation
	\begin{equation}
		\begin{aligned}
			\vec{A}_{r}H^{-1}\vec{F}_{r} = B_r
		\end{aligned}
		\label{eq:D2}.
	\end{equation}
	Note that the input force $\vec{F}_{r}$, as well as $\vec{A}_{r}, B_r$, are constants defined by the task to be learned, and the learning process only modifies elements of the Hessian $H$. For a system with $N$ physical degrees of freedom, we are allowed to modify the (at most) $\frac{1}{2}N(N+1)$ independent elements of the Hessian to find a solution to Eq.~\ref{eq:D2}.
	
	When training the system for $n_T$ simultaneous tasks, each defined by its own set of input $\vec{F}_{r}$ and constraints $\vec{A}_{r},B_r$, the system has to find a solution to $n_T$ independent linear equations at the same time. Learning attempts to modify the Hessian $H$ such that all of these equations are satisfied simultaneously. As shown, the system has at most $\frac{1}{2}N(N+1)$ free parameters to satisfy $n_T$ equations. A feasible solution exists only if the number of equations is at most equal to the number of free parameters. The network can thus learn a number of tasks which is at most equal to the number of independently trainable components of the Hessian. We thus expect a capacity proportional to the number of learning degrees of freedom, i.e., network weights, scaling at most quadratically with the system size and with a coefficient typically less than 1.
	This finite  capacity is visible in Fig.~\ref{fig:ManyTasks}a, where the error cannot be maintained at zero beyond $n_T\approx 13$, which also marks a turning point in the effective conductance of the network.
	
	%{\color{red} Below we suggest that the above analysis just for node networks.  We did not say that above.  If that is so, we should say that explicitly.  We seem to be contrasting with difference networks below, but I don't understand that difference... see my comment below.}
	Realistic physical networks are typically more constrained in the availability of adaptive (learning) degrees of freedom. In physical flow networks (e.g. vasculature) and mechanical networks (e.g. proteins), edges typically connect spatially neighboring nodes, so that the connectivity between nodes is short ranged, and the number of edges scales linearly with system size~\cite{rocks2019limits,ruiz2019tuning,stern2020continual}. Thus their learning capacity is linear in $N$. %{\color{red} VB: I am confused. The number of edges is still quadratic in the number of nodes, and hence the system size, as described above. So the learning capacity would appear to be quadratic again in $N$.  Please clarify}

	\section{Physical learning with noise}
	
	\begin{figure}
		\includegraphics[width=0.95\linewidth]{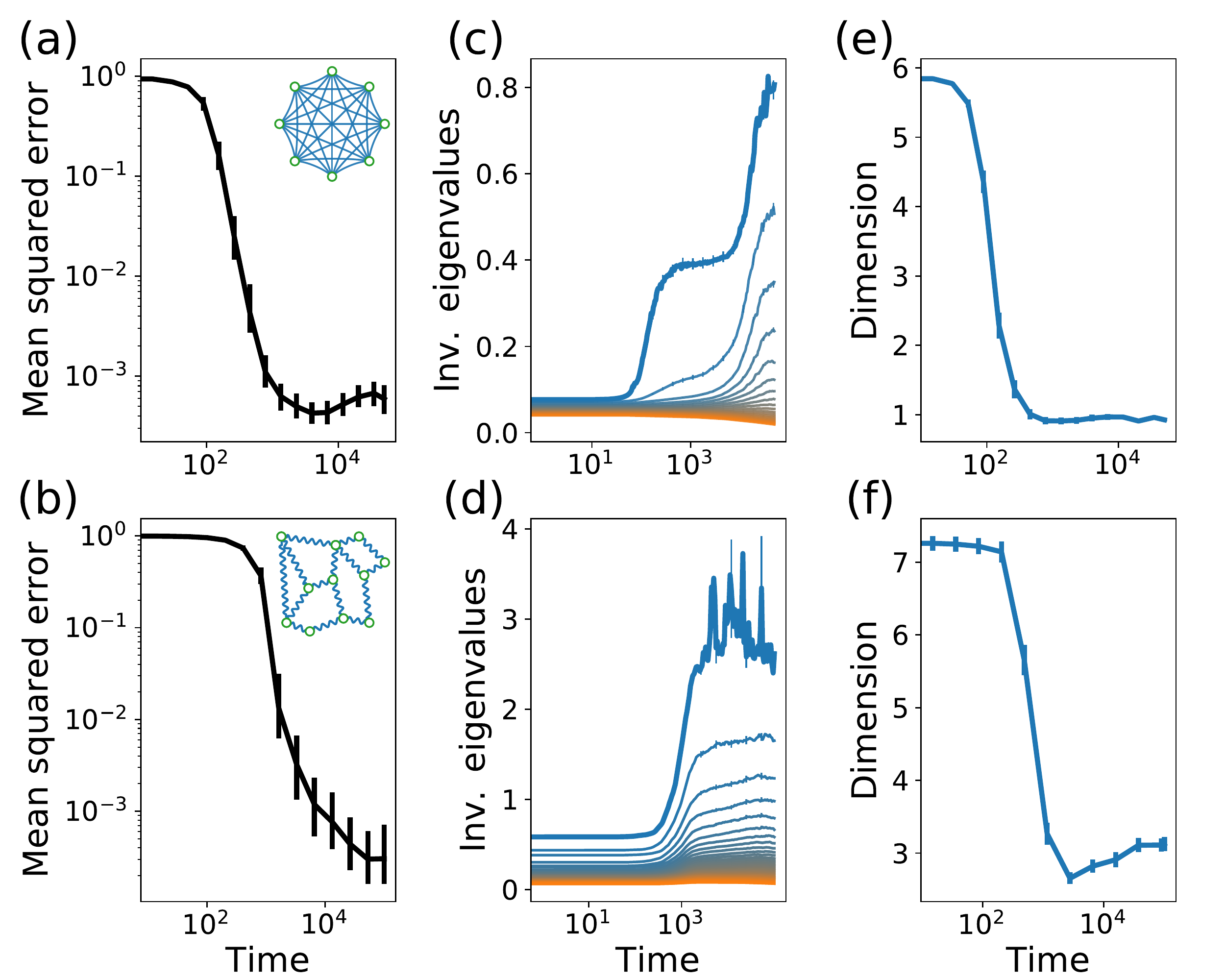}
		\caption{Noise limits the system's learning ability, but the physical effects of learning persist. (a-b) Training fully connected linear networks and mechanical spring networks for $5$ independent tasks with a noisy learning rule, we find that learning is still successful, yet is limited by an error floor. (c-d) However, the dynamics of the eigenspectrum are weakly affected by the noise, dominated by a reduction of the lowest lying eigenvalues (increaseing inverse eigenvalues). (e-f) Therefore, the physical properties of the networks trained with noise remain similar to the noiseless case. In particular, we observe that training sharply decreases the physical response dimension of the network. All results are averaged over $100$ sets of tasks.
			\label{fig:SI-Imperfections}}
	\end{figure}
	
	So far we discussed learning in pristine systems, where the learning rule is performed perfectly without any noise. In such cases, we found that physical learning can reduce errors arbitrarily (Fig.~\ref{fig:Errors}b) as long as the number of tasks is below the network capacity. However, it is  clear that real physical systems are prone to imperfections and  noise, and that these issues can limit  learning~\cite{dillavou2022demonstration}. In this appendix we discuss the effects of  noise on learning  and its physical effects on learning systems.

	A straightforward way to include imperfections in learning is to introduce additive Gaussian white noise in the learning rule (Eqs.~\ref{eq:2.a10},~\ref{eq:2.b4}). We add such noise, with a small amplitude compared to the learning rate $\alpha$, sampled from $\mathcal{N}(0,(5\cdot10^{-3} \alpha)^2)$, and train fully connected node networks and mechanical networks for five independent tasks (Fig.~\ref{fig:SI-Imperfections}).  The noise in the learning rule produces a floor in the error achievable by the network (Fig.~\ref{fig:SI-Imperfections}ab). 
	
	However, this noise does not strongly affect the physical dynamics of the Hessian and its eigenspectrum; compare Fig.~\ref{fig:SI-Imperfections}cd to Fig.~\ref{fig:FewTasks}. In this noisy case, learning still predominantly affects the lower eigenvalues, lowering them (increasing inverse eigenvalues) and aligning the associated eigenmodes with the task. We find that the noise tends to raise the upper eigenvalues, but as shown above, these typically have only minor influence on the physical responses of the system. As the eigenspectrum dynamics are largely unaffected by this noise, we can verify that the physical effects of learning discussed previously also persist, in particular the physical response dimension (Fig.~\ref{fig:SI-Imperfections}e,f).
	
	Besides noise in the learning degrees of freedom that would exist in any physical or biological realization, different learning protocols may introduce additional sources of noise. For example, consider stochastic gradient descent (SGD), where the system is trained at each iteration for a subset of the tasks. Although SGD introduces effective noise in the training dynamics, it is known to perform implicit regularization and improve generalization in computational machine learning~\cite{chaudhari2018stochastic}. Another plausible source of noise is the fact that in biological learning systems, the learning degrees of freedom are not synchronized, each evolving independently from the rest~\cite{kappel2015network}. We implement both SGD and the update desynchronization in our dynamics, letting the physical system train on one task and update $10\%$ of the learning degrees of freedom at every learning iteration. We note that in these simulations, we do not add white noise to the learning rule as done above, so that the desynchronized SGD learning dynamics are able to achieve perfect performance with vanishing error. While applying these protocols slows learning, we observe no qualitative difference in the physical properties of the trained system. These results suggest that the physical effects of learning in the linear regime that we described in this work are robust to noise that will likely exist in experimental realizations, such as recent experiments in learning resistor networks~\cite{dillavou2022demonstration, dillavou2023machine}.

	%\bibliographystyle{unsrt}
	%\bibliography{Citations.bib}

\end{document}